\documentclass[a4paper,fleqn,usenatbib]{mnras}

\usepackage[T1]{fontenc}
\usepackage{ae,aecompl}

\usepackage{graphicx}	
\usepackage{amsmath}	
\usepackage{amssymb}	
\usepackage{float}
\usepackage{natbib}
\usepackage{morefloats}
\usepackage{dcolumn}
\usepackage{amsmath}
\usepackage{latexsym}
\usepackage{longtable}
\usepackage{lipsum} 
\usepackage{gensymb}
\usepackage{adjustbox}
\usepackage{lscape}
\usepackage{hyperref}
\usepackage{rotating}

\newcommand{\kms}{\,km s$^{-1}$}	%
\newcommand{\Ha}{\,H$\alpha$}	%
\newcommand{\Hb}{\,H$\beta$}	%
\newcommand{\Msy}{\,M$_{\odot}$ yr$^{-1}$}
\newcommand{\Msun}{\,M$_{\odot}$}
\newcommand{\mhund}{\,mag (100d)$^{-1}$}

\hypersetup{draft}
\begin{document}
\title[Photometric and spectroscopic evolution of the peculiar Type IIn SN 2012ab]{Photometric and spectroscopic evolution of the peculiar Type IIn SN 2012ab}


\author[A.Gangopadhyay et al.]{Anjasha Gangopadhyay\thanks{e-mail :anjasha@aries.res.in, anjashagangopadhyay@gmail.com}$^{1,2}$, Massimo Turatto$^{3}$,  Stefano Benetti$^{3}$,
Kuntal Misra$^{1}$,
\newauthor
Brajesh Kumar$^{1,4}$, Enrico Cappellaro$^{3}$, Andrea Pastorello$^{3}$, Lina Tomasella$^{3}$,  
\newauthor
Sabrina Vanni$^{3}$,  Achille Fiore$^{3,5}$, A. Morales-Garoffolo$^{6}$, Nancy Elias-Rosa$^{9,10}$, 
\newauthor
Mridweeka Singh$^{1,2,7}$, Raya Dastidar$^{1,8}$, 
Paolo Ochner$^{5}$, Leonardo Tartaglia$^{9,10}$, 
\newauthor
Brijesh Kumar$^{1}$, Shashi Bhushan Pandey$^{1}$\\
\\
$^{1}$ Aryabhatta Research Institute of observational sciencES, Manora Peak, Nainital 263 001 India\\
$^{2}$ School of Studies in Physics and Astrophysics, Pandit Ravishankar Shukla University, Chattisgarh 492 010, India\\
$^{3}$ INAF Osservatorio Astronomico di Padova, Vicolo dell'Osservatotio 5, I-35122, Padova, Italy\\
$^{4}$ Indian Institute of Astrophysics, II-Block, Koramangala, Bengaluru 560 034, India \\
$^{5}$ Department of Physics and Astronomy Galileo Galilei, University of Padova, Vicolo dell'Osservatorio 3, I-35122, Padova, Italy\\
$^{6}$ Department of Applied Physics, University of Cadiz, Campus of Puerto Real, E-11510 Cadiz, Spain \\
$^{7}$ Korea Astronomy and Space Science Institute, 776 Daedeokdae-ro, Yuseong-gu, Daejeon 34055, Republic of Korea\\
$^{8}$ Department of Physics $\&$ Astrophysics, University of Delhi, Delhi-110 007\\
$^{9}$ Institut d’Estudis Espacials de Catalunya (IEEC), c/Gran Capit\'a 2-4, Edif. Nexus 201, 08034 Barcelona, Spain \\
$^{10}$ Institute of Space Sciences (ICE, CSIC), Campus UAB, Carrer de Can Magrans s/n, 08193 Barcelona, Spain
}

\date{Accepted XXX. Received YYY; in original form ZZZ}

\pubyear{2016}

\label{firstpage}
\pagerange{\pageref{firstpage}--\pageref{lastpage}}
\maketitle
\begin{abstract}
We present an extensive ($\sim$ 1200 d) photometric and spectroscopic monitoring  of the Type IIn supernova (SN) 2012ab. After a rapid initial rise leading to a bright maximum (M$_{R}$ = $-$19.39 mag), the light curves show a plateau lasting about 2 months followed by a steep decline up to about 100 d. Only in the $U$ band the decline is constant in the same interval. At later phases, the light curves remain flatter than the $^{56}$Co decline suggesting the increasing contribution of the interaction between SN ejecta with circumstellar material (CSM). Although heavily contaminated by emission lines of the host galaxy, the early spectral sequence (until 32 d) shows persistent narrow emissions, indicative of slow unshocked CSM, and the emergence of broad Balmer lines of hydrogen with P-Cygni profiles over a blue continuum, arising from a fast expanding SN ejecta. From about 2 months to $\sim$1200 d, the P-Cygni profiles are overcome by intermediate width emissions (FWHM $\sim 6000$ \kms), produced in the shocked region due to interaction. On the red wing a red bump appears after 76 d, likely a signature of the onset of interaction of the receding ejecta with the CSM. The presence of fast material both approaching and then receding is suggestive that we are observing the SN along the axis of a jet-like ejection in a cavity devoid of or uninterrupted by CSM in the innermost regions. 
\end{abstract}




\section{Introduction}
\label{sec:intro}
Type IIn supernovae \citep[SNe IIn, ][]{1990MNRAS.244..269S} constitute 6--9 per cent of core-collapse SNe (CCSNe) \citep{1997A&A...322..431C,2009MNRAS.395.1409S,2011MNRAS.418.1959S,2011MNRAS.412.1441L}, and their spectra are characterised by prominent lines of H and He I with emission components and composite profiles, although typically without showing broad P-Cygni absorption troughs. SNe IIn spectra show narrow-width (NW, $\sim100$ km sec$^{-1}$) components arising in the photoionised circumstellar medium (CSM) \citep{1994BAAS...26..791C,1998MNRAS.300L..17S}, along with intermediate-width (IW, $1000-3000$ km sec$^{-1}$) components due to either Thomson broadening of NW lines, or due to emission from gas shocked by the SN ejecta  \citep{1994MNRAS.268..173C,2001MNRAS.326.1448C,2009MNRAS.394...21D}. 
Some events show also very broad emission components arising from shocked ejecta \citep[e.g.][]{1993MNRAS.262..128T}.
Due to the dominance of strong CSM interaction, these 
SNe can be observed from X-ray to radio wavelengths, and may develop at later stages an IR excess resulting from re-processed radiation energy by pre-existing or newly formed dust  \citep[or a combination of the two processes,][]{2002ApJ...575.1007G,2009ApJ...691..650F}. SNe IIn are  luminous in comparison with other CCSNe.
Extended follow-up campaigns revealed a great observational diversity, probably reflecting different conditions of the CSM (e.g. structures, clumpiness, orientation) and SN parameters (e.g. energy, mass of the ejecta, mass loss history). This adds to great interest in studying these objects.
 
At the brighter end of SNe IIn luminosity class (M$_{v}\sim-22$ mag; \citealp{2012IAUS..279..253G}), objects like SN 2006gy \citep{2007ApJ...666.1116S,2007ApJ...659L..13O,2007ApJ...671L..17S,2009ApJ...691.1348A}, SN 2006tf \citep{2008ApJ...686..467S} and SN 2008iy \citep{2010MNRAS.404..305M} are found. \cite{2009ApJ...691.1348A} favoured CSM-interaction as source of the high luminosity, a scenario discussed by \cite{2013MNRAS.428.1020M} along with CSM being produced by pulsational instability of a zero-age main sequence 80--130 M$_{\odot}$ star. SN 2006tf and SN 2008iy are thought to arise from ejecta interaction with dense circumstellar shells ejected during the eruptions in Luminous Blue Variables (LBVs). At this luminosity we can also find the so-called SNe IIn-P, such as SNe 1994W, 2009kn and 2011ht \citep{2004MNRAS.352.1213C,2009MNRAS.394...21D,2012MNRAS.424..855K,2013MNRAS.431.2599M,2017MNRAS.472.3437G} that may originate from 8--10 M$_{\odot}$ stars undergoing core collapse as a result of electron capture after a brief phase of enhanced
mass loss, or from more massive (25 M$_{\odot}$) progenitors, which experience substantial fallback of the metal-rich radioactive material.
Some objects with similar moderate brightness (M$_{v}$ $\sim -$18 mag) may also result from intrinsically sub-luminous SN explosions interacting with CSM \citep{2004MNRAS.352.1213C,2016MNRAS.456.3296B} or objects like SN~1997cy that have narrow Balmer lines in addition to the broad features reminiscent to those observed in broad-lined Type Ic SNe \citep{2000ApJ...533..320G,2000ApJ...534L..57T,2003MNRAS.340..191R}. 

Narrow emission line spectra, somehow recalling SN IIn, are shown also by objects of  different nature like the so-called ``SN impostors'' \citep{2000PASP..112.1532V,2006MNRAS.369..390M} that are non-terminal eruptions of massive LBV stars. Sometimes these objects are followed by genuine core-collapse such as in SN 2009ip \citep[e.g.,][]{2013ApJ...767....1P,2014ApJ...787..163G,2014ApJ...780...21M,2014MNRAS.438.1191S,2015MNRAS.453.3886F,2015AJ....149....9M,2013MNRAS.430.1801M,2015hst..prop14150F,2017MNRAS.472.3437G}.

The current understanding of the progenitor of SNe IIn is not fully established. Since SNe IIn have dense CSM draining kinetic energy from the blast wave, the most luminous SNe requires high mass loss rates of the order of 0.01 M$_{\odot}$  yr$^{-1}$ \citep{2004MNRAS.352.1213C, 2007ApJ...671L..17S}. Such extreme environments point to episodic mass loss rates reminiscent of the eruptions as seen in $\eta-$Carinae and other LBVs having mass loss rates upto $\leq1$ M$_{\odot}$ yr$^{-1}$ far exceeding the limits of line-driven winds \citep{2006ApJ...645L..45S}. The detection of their LBV like progenitors was claimed in the past for SNe 2005gl, 2009ip and 2010jl \citep{2009Natur.458..865G,2013MNRAS.430.1801M,2011ApJ...732...63S,2016MNRAS.463.2904S}. This opens the way to consistently large mass loss rates with respect to conventional core collapse progenitors, ranging from 10$^{-4}$ M$_{\odot}$ yr$^{-1}$ for Red supergiants (RSGs), 10$^{-5}$ M$_{\odot}$ yr$^{-1}$  for Yellow supergiants (YSGs) \citep{2017MNRAS.471.4047A} and $\sim$10$^{-4}$ M$_{\odot}$ yr$^{-1}$ for Wolf-Rayet stars \citep{2012ApJ...744...10K}.

In a few cases, it is interesting to note that SNe with narrow line spectra, such as SNe 2002ic and 2005gj have been also associated with thermonuclear explosions \citep{2003Natur.424..651H,2006ApJ...650..510A}, although \cite{2006ApJ...653L.129B} argued alternatively suggested the core-collapse of a stripped-envelope star interacting with a dense CSM as a more plausible explanation. However the Type Ia PTF~11kx \citep{2012Sci...337..942D} (and more recently SN~2015cp, \citealp{2019ApJ...871...62G} and ASASSN-18tb, \citealp{2019arXiv190202251K}) showed convincingly that SN Ia can also be surrounded by multiple shells of CSM. 
A class of transients, called Tidal Disruption Events (TDE), have been observed to occur in the nuclei of galaxies showing well defined characteristics, like blue continuum and strong lines of He II and H \citep{2011ApJ...741...73V,2012Natur.485..217G}. These have been interpreted as the result of the tidal disruption of stars by massive black hole in the galaxy nuclei \cite[e.g.][]{1988Natur.333..523R}. 
More recently, population of highly energetic transients in the centres of active galaxies has been identified. Important members of this class are PS1-10adi, SN 2006gy, CSS100217:102913
+404220, ASASSN-15lh \citep{2017NatAs...1..865K, 2007ApJ...666.1116S, 2011ApJ...735..106D,2016Sci...351..257D}.
The origin of these bright nuclear transients, that in the past was overlooked because of incorrectly associated with black hole activity, is debated. 
PS1-10adi is the most energetic and one of the best studied nuclear SN. \cite{2017NatAs...1..865K} suggest that the high energy originates from shock interaction between expanding gas ejecta and with quantities of surrounding dense matter. The expanding material mostly arises from a star that has been tidally disrupted by the central black hole or a SN. However, such bright transients may also originate from reprocessed emission from an active galactic nuclei.

Recently, \cite{2018MNRAS.475.1104B} reported the photometric, spectroscopic and spectropolarimetric study of SN 2012ab which exploded close to the nucleus in the central region of the host galaxy. 
SN 2012ab was a high luminosity (M$\sim$ $-19.4$ mag) SN IIn showing strong intermediate-width features due to CSM interaction, along with absence of broad absorption features. Spectropolarimetry suggested strong asymmetry in the CSM structure, with the progenitor being in part an eccentric binary system undergoing eruptive mass loss stages.
In this paper, we extend the study of SN 2012ab with multiband photometry and extensive spectroscopic coverage until $\sim$ 1200 d post maximum. We provide information on the SN discovery and its host galaxy in Section \ref{sec:galaxy}. 
Section \ref{sec:data} describes the data acquisition and reduction procedures. In Section \ref{sec:lc}, we study the multi-band light curves, and compare the colours, absolute light curves and bolometric evolution with those of other SNe IIn. The spectroscopic evolution is described in Section \ref{sec:spec}.
In Section \ref{sec:ha_profile}, we analyse in detail the evolution of the \Ha\/ line profile. Finally in Section \ref{sec:discussion}, we estimate the mass-loss rates and provide a plausible progenitor scenario for SN 2012ab. A summary is given in Section \ref{sec:conclusions}.

\section{SN 2012ab and its host galaxy}
\label{sec:galaxy}
SN 2012ab was discovered by J. Vinko and the ROTSE collaboration, at $\sim$ 15.8 mag (on 2012 January 31.35 UT (MJD = 55957.86) with the 0.45m ROTSE-IIIb telescope at the McDonald Observatory. The object was located at R.A.=12$^{h}$22$^{m}$47.6$^{s}$, Decl.=+05$^{\circ}$36$^{'}$25.0$^{''}$ (J2000.0), in the proximity of the center of the host galaxy SDSS J122247.61+053624.2, at a redshift of 0.018 \citep{2012CBET.3022....1V}. The  details of the host galaxy are given in Table \ref{tab:galaxydetails}. No source was detected at the SN location before 2012 January 29.28, down to a limiting magnitude $\sim$ 18.7 mag.

\cite{2000ApJ...529..786M} simulated a velocity field model to estimate the recessional velocity of the host galaxy recession. The velocity model corrected for the influences of Virgo cluster, the Great Attractor and the Shapley supercluster gives 6010 $\pm$ 40 km sec$^{-1}$. Assuming H$_{0}$ = 73 $\pm$ 5 km sec$^{-1}$ Mpc$^{-1}$, this corresponds to a luminosity distance of 82.3 $\pm$ 5.8 Mpc.  The same value is also used by \cite{2018MNRAS.475.1104B}. 

In order to estimate the contribution of the host galaxy to the total reddening, we  measured the equivalent width of the NaID 5890, 5896 \AA~ doublet at the host galaxy redshift in a couple of good resolution spectra (2012 May 26, 2012 June 08 and 2012 June 27; resolutions are listed in Table \ref{tab:spec_obs}). A number of relations have been proposed to correlate the equivalent width of the NaID lines to the dust extinction \citep{1994AJ....107.1022R, 1997A&A...318..269M, 2003fthp.conf..200T,  2012MNRAS.426.1465P}. The average value of equivalent widths for the D1 and D2 lines are estimated to be 0.24 $\pm$ 0.03 \AA~ and 0.30 $\pm$ 0.05 \AA, respectively. Using the \cite{2012MNRAS.426.1465P} relation and the equivalent widths, we estimate the host galaxy extinction along the line of sight to be A$_{V}$ = 0.19 $\pm$ 0.10 mag, assuming A$_{V}$ = $3.08 \times E(B-V)$ \citep{1992ApJ...395..130P}. We also estimated the equivalent widths of NaID using the formulations by \cite{1994AJ....107.1022R, 1997A&A...318..269M, 2003fthp.conf..200T,  2012MNRAS.426.1465P} which yielded the results well within the error bars of our obtained values. We also adopt the galactic reddening \citep{2011ApJ...737..103S} in the direction of SN 2012ab. Thus, the total reddening value is  A$_{V}$ = 0.24 mag for $E(B-V)$ = 0.079 mag, which is in good agreement with \cite{2018MNRAS.475.1104B}.

Several diagnostics are used to estimate the host galaxy metallicity \citep{1991ApJ...380..140M, 2002ApJS..142...35K,2004MNRAS.348L..59P,2005ApJ...631..231P}. We used the SDSS spectrum of the host galaxy taken on  2008 February 13\footnote{\it http://skyserver.sdss.org/dr14/en/get
/SpecById.ashx?id=3242705331930818560} in which both [O III] 5007 \AA~ and [N II] 6583 \AA~ lines are clearly visible. Using the O3N2 index calibration by \cite{2004MNRAS.348L..59P} we estimate the host galaxy metallicity to be 12 + log(O/H)$\sim 8.5\pm0.5$ dex. As a comparison, \citet{2001ApJ...558..830A}, \citet{2009ARA&A..47..481A} and \citet{2011SoPh..268..255C} estimated the solar metallicity  to be 8.69 $\pm$ 0.05 dex, 8.69 dex and 8.76 $\pm$ 0.07 dex, respectively. Thus, the host galaxy metallicity is nearly solar.

\begin{table}
\caption {General information about SN 2012ab and the host galaxy. The unfiltered magnitude of the SN is scaled to the R band photometry. \label{tab:galaxydetails}}
\begin{center}
\begin{tabular}{cc} \hline\hline
\multicolumn{2}{c}{\bfseries SN 2012ab} \\ \hline
{\bfseries Discovery Date} & 2012 Jan. 31.35 (MJD 55957.86)\\
{\bfseries Explosion Date} & 2012 Jan. $29.3\pm1$ (MJD 55955.3)\\
{\bfseries SN Type} & IIn\\
{\bfseries RA (J2000.0)} & 12$^{h}$ 22$^{m}$ 47.60$^{s}$ \\
{\bfseries Dec (J2000.0)}  &  +05$^{\circ}$ 36$^{'}$ 25$^{''}$.0\\
{\bfseries Discovery Magnitude} &  16.3 (Unfiltered)\\
{\bfseries E(B-V)$_{tot}$} & 0.079 mag \\
\multicolumn{2}{c}{{\bfseries HOST GALAXY}}\\ \hline
{\bfseries Galaxy Names} & SDSS J122247.61+053624.2 \\
{\bfseries } & 2MASXJ12224762+0536247 \\
{\bfseries } & LEDA 1286171 \\
{\bfseries Morphology Type} & Spiral \\
{\bfseries RA (J2000.0)} & 12$^{h}$ 22$^{m}$ 47.61$^{s}$ \\
{\bfseries Dec (J2000.0)}  & +05$^{\circ}$ 36$^{'}$ 24.3$^{''}$ \\
{\bfseries redshift (z)} & 0.018 \\
{\bfseries v$_{Virgo+Shapley}$ (km s$^{-1}$)} & 6010$\pm$40\\
{\bfseries m$_B$ } & 16.60 mag \\
{\bfseries M$_B$ } & $-$17.84 mag \\
{\bfseries 12+log(O/H)}  & 8.5$\pm$0.5 dex\\
{\bfseries d (Mpc)} & 82.30$\pm$5.8\\ 
{\bfseries $\mu$ (mag)} &34.57\\\hline
\end{tabular}  
\end{center}
\end{table}

\subsection{SN Location}
\label{sec:SNloc}
To determine the exact location of the SN in the host galaxy, we  performed astrometric calibration of the SDSS DR14 image of the host galaxy taken in November 2001, and the galaxy-subtracted SN image taken in September 2012 with 1.0m ST, ARIES using {\it astrometry.net} \citep{2008AJ....136.1490B}. We use the same technique as \cite{2018MNRAS.475.1104B} for determining the SN and the host galaxy nucleus location by performing radial profile fits to the Moffat distribution. The uncertainty in the position is determined by refitting the centroid 100 times and quoting the rms of the covered measurements as the uncertainty. The precise position of the SN is $\alpha$ = 12$^{h}$22$^{m}$47.60$^{s}$, $\delta$ = +05$^{\circ}$36$^{'}$25$^{''}.00$ and the host galaxy core is at $\alpha$ = 12$^{h}$22$^{m}$47.60$^{s}$,  $\delta$ = +05$^{\circ}$36$^{'}$24$^{''}.30$ (J2000). The estimated offsets in $\alpha$ and $\delta$ values are 
0.01$^{s}$ $\pm$ 0.01$^{s}$  and 0.70$^{''}$ $\pm$ 0.09$^{''}$, 
respectively, in close agreement with those found by \cite{2018MNRAS.475.1104B}. So, we conclude that the position of SN 2012ab is displaced by about 280 pc; allowed for projection effects; away from the host galaxy center and not coincident with the nucleus. This formally rules out that SN 2012ab is not strictly a nuclear transient, such as an AGN flare or a TDE.

\begin{figure*}
	\begin{center}
		\hspace{-1.0cm}
		\includegraphics[scale=0.21]{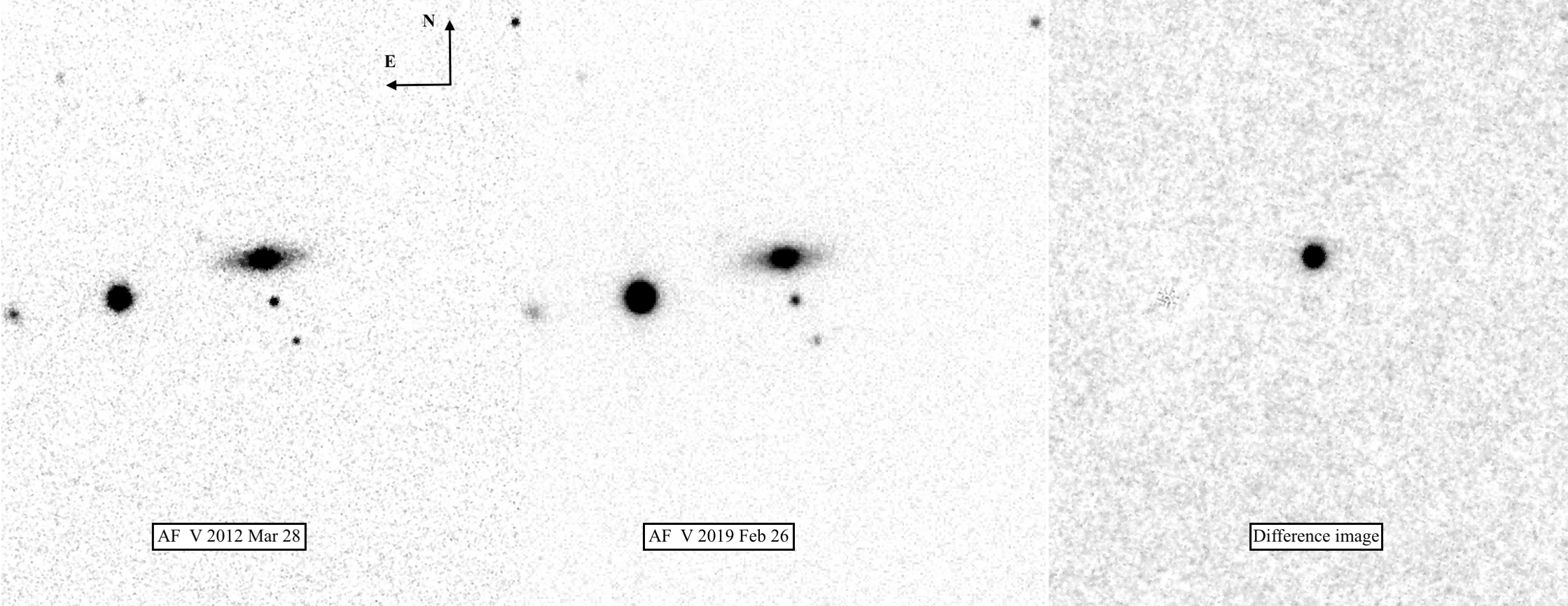}
	\end{center}
	\caption{The left panel shows the image of SN 2012ab taken on 2012 March 28 with 1.8m Mt. Ekar telescope. The central panel shows the image of the parent galaxy taken $\sim$ 7 years after the explosion with the same telescope. The right panel shows the template subtracted image of SN 2012ab (cfr. Fig.~\ref{fig:calibimage}).}
	\label{fig:subplot}
\end{figure*}

\begin{figure*}
	\begin{center}
		\hspace{-1.0cm}
		\includegraphics[scale=0.5]{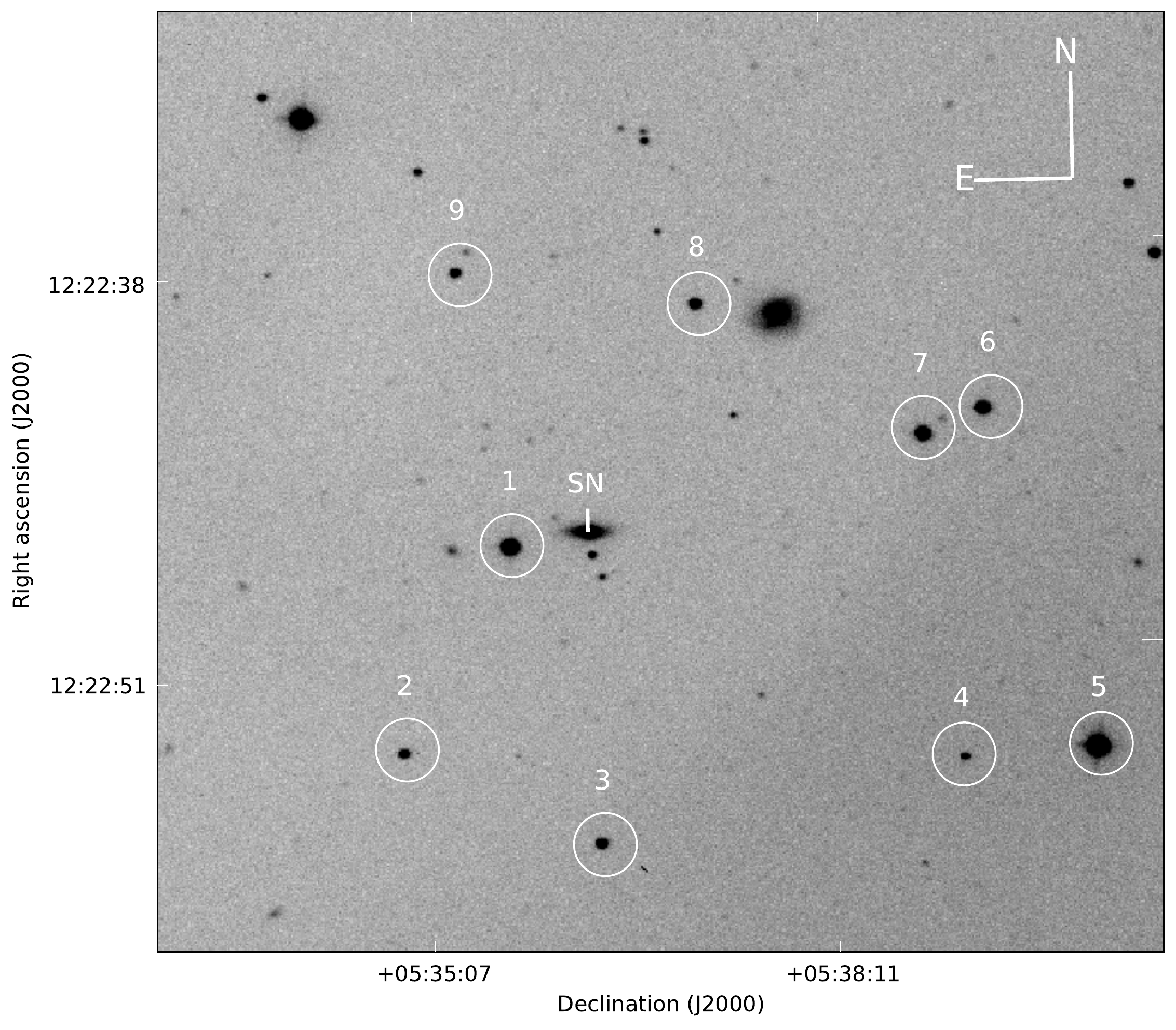}
	\end{center}
	\caption{SN 2012ab and the local standard stars in the $8\times8$ arcmin$^2$ field of SDSS J122247.61+053624.2 in a $R$-band image obtained on 2012 February 09 with the 1.04m ST. The magnitude of the standard stars are given in Table \ref{tab:optical_observations}.}
	\label{fig:calibimage}
\end{figure*}

\begin{center}
\begin{table*}
\caption{Apparent magnitudes of SN 2012ab in optical bands.}  		  
\begin{tabular}{c c c c c c c c c c}
\hline \hline
Date        & Phase$^{\dagger}$  & MJD              & $U$               & $B$             & $V$           & $R$          &  $I$         & Instrument\\    
(dd/mm/yy)  & (d)  &                  & (mag)             & (mag)           & (mag)         & (mag)        & (mag)        &            \\
\hline
 20/02/12 & 22.62  & 55977.82 & 15.10 $\pm$  0.06  & 15.98 $\pm$   0.027 & 15.77 $\pm$  0.02 & 15.50 $\pm$  0.01  & 15.43 $\pm$  0.02 & 1   \\    
 21/02/12 & 23.57  & 55978.77 & ---                & ---                  & ---              & 15.31 $\pm$  0.03  & 15.35 $\pm$  0.04 & 1 \\ 
 22/02/12 & 24.57  & 55979.77 & 15.05 $\pm$  0.13  & 15.93 $\pm$   0.02 & 15.69 $\pm$   0.03 & ---                & ---                 & 1 \\ 
 05/03/12 & 36.61  & 55991.81 & 15.55 $\pm$  0.08  & 16.44 $\pm$   0.04 & ---                & 15.58 $\pm$  0.02  & 15.48 $\pm$  0.03 & 1 \\ 
 10/03/12 & 40.96  & 55996.16 & ---                & 16.40 $\pm$   0.05 & 15.91 $\pm$   0.05 & 15.57 $\pm$  0.08  & ---                 & 2 \\ 
 12/03/12 & 42.91  & 55998.11 & ---                & 16.45 $\pm$   0.01 & 15.98 $\pm$   0.01 & 15.59 $\pm$  0.01  & 15.58 $\pm$  0.01 & 3 \\ 
 12/03/12 & 43.79  & 55998.99 & ---                & 16.58 $\pm$   0.04 & 16.13 $\pm$   0.04 & 15.67 $\pm$  0.02  & 15.61 $\pm$  0.03 & 1 \\ 
 13/03/12 & 44.76  & 55999.96 & ---                & ---                & 16.02 $\pm$   0.02 & 15.62 $\pm$  0.03  & 15.55 $\pm$  0.02 & 2 \\ 
 14/03/12 & 44.87  & 56000.07 & 16.00 $\pm$  0.02  & 16.53 $\pm$   0.02 & ---                  & ---                 & ---                 & 4 \\ 
 14/03/12 & 45.66  & 56000.86 & 15.93 $\pm$  0.06  & 16.52 $\pm$   0.03 & 16.05 $\pm$   0.03 & 15.63 $\pm$  0.04  & 15.63 $\pm$  0.04 & 1 \\ 
 17/03/12 & 47.97  & 56003.17 & ---                & 16.63 $\pm$   0.02 & 16.09 $\pm$   0.02 & 15.61 $\pm$  0.02  & 15.52 $\pm$  0.03 & 4 \\ 
 24/03/12 & 55.74  & 56010.94 & 16.05 $\pm$  0.03  & 16.51 $\pm$   0.02 & 16.08 $\pm$   0.02 & 15.62 $\pm$  0.02  & 15.48 $\pm$  0.01 & 3 \\ 
 26/03/12 & 57.64  & 56012.84 & 16.24 $\pm$  0.05  & 16.56 $\pm$   0.03 & 16.17 $\pm$   0.03 & 15.67 $\pm$  0.02  & 15.67 $\pm$  0.04 & 1 \\ 
 28/03/12 & 59.71  & 56014.91 & ---                & 16.58 $\pm$   0.03 & 16.12 $\pm$   0.04 & 15.77 $\pm$  0.06  & 15.63 $\pm$  0.03 & 2 \\ 
 28/03/12 & 59.73  & 56014.93 & 16.16 $\pm$  0.05  & 16.56 $\pm$   0.03 & 16.24 $\pm$   0.03 & 15.85 $\pm$  0.03  & 15.77 $\pm$  0.03 & 4 \\ 
 28/03/12 & 59.75  & 56014.95 & 16.06 $\pm$  0.03  & 16.72 $\pm$   0.01 & 16.13 $\pm$   0.02 & 15.77 $\pm$  0.02  & 15.63 $\pm$  0.02 & 3 \\ 
 01/04/12 & 62.82  & 56018.02 & ---                & 16.73 $\pm$   0.01 & 16.24 $\pm$   0.02 & 15.91 $\pm$  0.04  & 15.65 $\pm$  0.05 & 2 \\ 
 01/04/12 & 63.53  & 56018.73 & 16.30 $\pm$  0.14  & 16.74 $\pm$   0.02 & ---                & 15.87 $\pm$  0.03  & 15.84 $\pm$  0.06 & 1 \\
 03/04/12 & 65.56  & 56020.76 & 16.25 $\pm$  0.09  & 16.82 $\pm$   0.03 & ---                & 15.96 $\pm$  0.03  & 15.89 $\pm$  0.04 & 1 \\ 
 04/04/12 & 66.57  & 56021.77 & ---                & ---                & 16.42 $\pm$   0.03 & ---                 & ---                 & 1 \\ 
 08/04/12 & 70.40  & 56025.60 & 16.39 $\pm$  0.10  & ---                & 16.64 $\pm$   0.04 & 16.16 $\pm$  0.02  & 16.09 $\pm$  0.04 & 1 \\
 13/04/12 & 75.59  & 56030.79 & 16.76 $\pm$  0.07  & 17.15 $\pm$   0.04 & 16.78 $\pm$   0.03 & 16.21 $\pm$  0.04  & 16.27 $\pm$  0.04 & 1 \\ 
 14/04/12 & 75.96  & 56031.16 & 16.31 $\pm$  0.02  & 17.04 $\pm$   0.05 & 16.84 $\pm$   0.06 & 16.13 $\pm$  0.03  & 16.38 $\pm$  0.03 & 5 \\ 
 19/04/12 & 81.50  & 56036.70 & 17.04 $\pm$  0.10  & 17.42 $\pm$   0.05 & 17.12 $\pm$   0.03 & 16.43 $\pm$  0.03  & 16.49 $\pm$  0.05 & 1 \\ 
 24/04/12 & 86.63  & 56041.83 & 17.21 $\pm$  0.13  & 17.58 $\pm$   0.05 & 17.05 $\pm$   0.05 & 16.48 $\pm$  0.03  & 16.47 $\pm$  0.04 & 1 \\ 
 24/04/12 & 86.78  & 56041.98 & ---                & 17.38 $\pm$   0.03 & 16.89 $\pm$   0.03 & 16.63 $\pm$  0.05  & 16.56 $\pm$  0.04 & 2 \\ 
 27/04/12 & 89.69  & 56044.89 & ---                & 17.63 $\pm$   0.03 & ---                & 16.41 $\pm$  0.06  & 16.53 $\pm$  0.04 & 2 \\
 29/04/12 & 91.40  & 56046.60 & 17.14 $\pm$  0.13  & 17.83 $\pm$   0.09 & 17.24 $\pm$   0.06 & 16.59 $\pm$  0.04  & 16.63 $\pm$  0.05 & 1 \\ 
 30/04/12 & 92.79  & 56047.99 & ---                & 17.92 $\pm$   0.03 & 17.23 $\pm$   0.03 & 16.88 $\pm$  0.08  & 16.88 $\pm$  0.03 & 6   \\
 07/05/12 & 99.52  & 56054.72 & ---                & 18.02 $\pm$   0.13 & 17.39 $\pm$   0.18 & 16.84 $\pm$  0.07  & 17.07 $\pm$  0.17 & 1    \\ 
 13/05/12 & 105.52 & 56060.72 & 17.35 $\pm$  0.12  & 18.13 $\pm$   0.11 & 17.39 $\pm$   0.07 & 16.74 $\pm$  0.03  & 16.87 $\pm$  0.06 & 1    \\ 
 15/05/12 & 107.76 & 56062.96 & 17.30 $\pm$  0.10  & 18.08 $\pm$   0.02 & ---                & 16.84 $\pm$  0.01  & 17.04 $\pm$  0.09 & 3    \\ 
 17/05/12 & 109.49 & 56064.69 & 17.43 $\pm$  0.09  & 18.07 $\pm$   0.04 & 17.54 $\pm$   0.06 & 16.80 $\pm$  0.03  & 17.09 $\pm$  0.08 & 1    \\ 
 26/05/12 & 118.54 & 56073.74 & 17.86 $\pm$  0.16  & 18.24 $\pm$   0.19 & 17.33 $\pm$   0.10 & 16.91 $\pm$  0.05  & 17.15 $\pm$  0.08 & 1    \\ 
 26/05/12 & 118.73 & 56073.93 & 17.89 $\pm$  0.02  & 18.34 $\pm$   0.02 & 17.55 $\pm$   0.02 & 16.99 $\pm$  0.02  & 17.23 $\pm$  0.05 & 7    \\ 
 08/06/12 & 130.80  & 56086.00 & 17.88 $\pm$ 0.02  & 18.37 $\pm$   0.02 & 17.61 $\pm$   0.03 & 17.06 $\pm$  0.02  & 17.31 $\pm$  0.03 & 8    \\ 
 09/06/12 & 132.53 & 56087.73 & 18.05 $\pm$  0.37  & 18.43 $\pm$   0.08 & 17.50 $\pm$   0.08 & 17.07 $\pm$  0.05  & 17.32 $\pm$  0.12 & 1    \\ 
 25/06/12 & 148.46 & 56103.66 & ---                & 18.35 $\pm$   0.15 & 17.74 $\pm$   0.10 & 17.11 $\pm$  0.07  & 17.57 $\pm$  0.19 & 1    \\ 
 27/06/12 & 150.65 & 56105.85 & ---                & ---                & 17.74 $\pm$   0.08 & ---                 & ---                 & 4 \\
 07/07/12 & 160.75 & 56115.95 & 18.27 $\pm$  0.08  & 18.71 $\pm$   0.05 & 17.89 $\pm$   0.05 & 17.19 $\pm$  0.04  & 17.68  $\pm$  0.06 & 6   \\ 
 18/07/12 & 171.64 & 56126.84 & ---                & ---                & 18.03 $\pm$   0.09 & 17.19 $\pm$  0.06  & 17.74 $\pm$  0.09 & 4   \\ 
 28/11/12 & 304.75 & 56259.95 & ---                & 19.05 $\pm$   0.15 & 18.46 $\pm$   0.06 & 17.73 $\pm$  0.04  & 18.07 $\pm$  0.24 & 1   \\
 10/12/12 & 315.78 & 56270.98 & 18.43 $\pm$  0.21  & 18.96 $\pm$   0.08 & 18.53 $\pm$   0.06 & 17.62 $\pm$  0.09  & 18.16 $\pm$  0.15 & 1   \\ 
 13/01/13 & 350.71 & 56305.91 & ---                & ---                & 18.62 $\pm$   0.08 & 17.81 $\pm$  0.05  & 18.54 $\pm$  0.16 & 1   \\
 14/01/13 & 351.78 & 56306.98 & 18.72 $\pm$  0.16  & 19.14 $\pm$  0.09  & ---                  & ---              & ---                 & 1   \\
 03/02/13 & 371.66 & 56326.86 & ---                & 19.59 $\pm$  0.15  & 18.90 $\pm$   0.15 & 18.03 $\pm$  0.08  & 18.61 $\pm$  0.24 & 1   \\ 
 09/02/13 & 376.94 & 56332.14 & 18.95 $\pm$  0.17  & 19.75 $\pm$  0.08  & 18.82 $\pm$   0.04 & 18.08 $\pm$  0.03  & 18.77 $\pm$  0.11 & 4   \\
 21/02/13 & 389.76 & 56344.96 & ---                & 19.63 $\pm$  0.13  & 18.88 $\pm$   0.09 & 18.05 $\pm$  0.07  & 18.72 $\pm$  0.14 & 1   \\ 
 16/03/13 & 411.95 & 56367.15 & ---                & >19.94 $\pm$ 0.20  & 19.29 $\pm$   0.07 & 18.24 $\pm$  0.06  & 18.75 $\pm$  0.11 & 4   \\
 11/04/13 & 438.61 & 56393.81 & ---                & 20.00  $\pm$ 0.14  & 19.41 $\pm$   0.11 & 18.48 $\pm$  0.06  & 19.16 $\pm$  0.21 & 1  \\
 25/04/14 & 816.79 & 56771.99 & ---                & ---                & 19.55 $\pm$   0.10 & 18.95 $\pm$  0.43  & 19.62 $\pm$  0.25  & 2 \\
 12/02/15 & 1109.96& 57065.16 & ---                &>20.39 $\pm$  0.20  & 19.99 $\pm$   0.19 & 19.65 $\pm$  0.19  & 20.22 $\pm$  0.41 & 4   \\
 11/06/15 & 1229.72& 57184.92 & ---                &>20.44 $\pm$  0.20  & >20.18 $\pm$ 0.20 & 19.93 $\pm$ 0.23  & > 20.66 $\pm$  0.20 & 4 \\
\hline          		                    
\end{tabular}
\label{tab:observation_log}     
\newline
${\dagger}$ With respect to explosion (MJD=55955.20$\pm$0.5)
\newline
1 ARIES 1.04m ST+2kx2k ccd; 2 Ekar Schmidt 67/92cm; 3 Calar Alto 2.2m+CAFOS; 4 Ekar 1.82m+AFOSC; 5 ESO NTT 3.58m+EFOSC2; 6 NOT 2.54m+ALFOSC; 7 WHT 4.20m+ACAM; 8 TNG 3.58m+LRS
\end{table*}
\end{center}

\begin{table*}
\caption{Apparent magnitudes of SN 2012ab in $JHK$ bands. The value reported for the {\it K} band of NICS mounted at the 3.58m Telescopio Nazionale 
Galileo (TNG), was obtained with the $K^{'}$ filter. \label{photo_NIR}}
\begin{center}
\begin{tabular}{c c c c c c c}
\hline
{\bfseries Data} & {\bfseries Phase$^{\dagger}$} & {\bfseries MJD} &{\bfseries $J$} & {\bfseries $H$} & 
{\bfseries $K$} &{\bfseries Instrument}\\
{\bfseries (dd/mm/yy)}& (d) & & & & &  \\
\hline 
\hline
15/03/12 & 46.48 & 56001.68 & 14.97 $\pm$ 0.06 &14.59 $\pm$ 0.08&14.29  $\pm$ 0.11 & 9\\
04/06/12 & 127.33 & 56082.53 & 15.62 $\pm$ 0.04 & 15.14 $\pm$ 0.04 &15.03 $\pm$ 0.07 & 10\\
04/07/12 & 158.22 & 56113.42 & 15.88  $\pm$ 0.03 & 15.18 $\pm$ 0.05&14.99 $\pm$ 0.03 & 11 \\
\hline
\label{tab:IR_mag} 
\end{tabular}
\newline
$\dagger$ With respect to the explosion (MJD=55955.20$\pm$0.5)
\newline
Note: 9 = ES0 NTT 3.58m +SOFI; 10 = TNG 3.58m+NICS; 11 = NOT 2.54m+NOTCam.
\end{center}
\end{table*}

\section{Data Acquisition and Reduction}
\label{sec:data}
\subsection{Photometric Observations}
\label{sec:phot}
Photometric follow-up observations of SN 2012ab were carried out for about 3 years, from 2012 February 20 to 2015 June 11, using a large number of ground-based optical telescopes listed in Tables ~\ref{tab:observation_log} and \ref{tab:IR_mag}.

Imaging was done using Johnson-Cousins-Bessell {\it U, B, V, R, I} filters. Basic pre-processing and image reduction were done using IRAF\footnote{Image Reduction and Analysis Facility} as outlined in \cite{2018MNRAS.476.3611G}. SN 2012ab is embedded near the center of the host galaxy, therefore template subtraction has revealed it to be necessary for accurate measurements at all epochs. As templates, we used a set of deep images obtained on 2019 February 26 with the 1.82m Copernico Telescope at Ekar-Asiago equipped with AFOSC, when the SN was well beyond the limit of detection. 
SN and template images were aligned, scaled and subtracted using the SNOoPY\footnote{Cappellaro, E. (2014). SNOoPY: a package for SN photometry, http://sngroup.oapd.inaf.it/snoopy.html} package which also performs point-spread-function measurements of the SN on the galaxy subtracted images.
The instrumental magnitudes were then calibrated using colour terms previously determined  with the same instruments on a number of standard fields \citep{1992AJ....104..340L}, while zero-points were determined for each night with reference to a previously calibrated local sequence of stars (cfr. Fig.~\ref{fig:calibimage}, Table \ref{tab:optical_observations}).
The magnitudes of the stars of the local sequence were derived on a number of photometric nights. These resulted in good agreement ($\Delta(\rm m)< 0.03$ mag) with the  magnitudes derived transforming the SLOAN A$_B$ magnitudes to the Vega {\it UBVRI} ones with the transformations \citep{2008AJ....135..264C}. Only the brightest star n.1 showed large deviations possibly because it is saturated in the SLOAN survey.
The errors due to the calibration and the photometric measurements were added in quadrature to estimate the final error of the SN magnitudes.

ROTSE-IIIb obtained an intensive monitoring coverage of the host galaxy field before and after the SN discovery. The original unfiltered images were calibrated to $R$-band via USNO-B1.0 photometric calibration \citep[Table~1 of][]{2018MNRAS.475.1104B}.
We used our local sequence to check this calibration finding a fair agreement ($<\Delta(\rm m)> \simeq 0.02$ mag).  Therefore, in order to describe the SN behaviour shortly after the shock-breakout, a period in which we are short of data, we use the early $R$-band photometry from \cite{2018MNRAS.475.1104B} (cfr. Fig.~\ref{fig:lc}).

Near-infrared (near-IR) photometry ({\it JHK}) was obtained with different telescopes. The reduction of IR images needed a few additional steps. After the flat field correction, sky images were obtained median-combining several science frames. Sky images were then subtracted to individual science images in order to remove the contribution of the bright near-IR background. We remark that the IR images are obtained from PSF-fitting and not template subtraction. Finally, science frames were aligned and combined in order to improve the signal-to-noise (S/N). Photometric calibration was achieved relative to the 2MASS\footnote{Two Micron All Sky Survey.} magnitudes of the same local sequence stars used for the calibration of the optical photometry.
The SN magnitude determinations in optical and IR bands are reported in Table~ \ref{tab:observation_log} and Table ~\ref{tab:IR_mag}, respectively.

\subsection{Spectroscopic observations}
\label{spec}

We obtained low and medium resolution optical and NIR spectra of SN 2012ab at 26 epochs (from 2012 February 7 to 2015 February 27), as tabulated in Table \ref{tab:spec_obs}. 
Unfortunately our spectroscopic campaign missed the earliest epochs after the discovery of SN 2012ab and for the description of the early phases we rely on the Hobby-Eberly Telescope (HET) spectra presented in \citet{2018MNRAS.475.1104B}), as reported in Table ~\ref{tab:spec_obs}. The 2-d spectra were pre-processed using standard tasks in {\textsc IRAF}, as well as for the subsequent extraction and calibration of the 1-d spectra. HeNe, HeAr, HgCdNe  arc lamps were used for wavelength calibration, whose accuracy was checked using night-sky emission lines. When appropriate, minor rigid wavelength shifts were applied. Spectroscopic standards were observed for correcting the instrumental response, and for flux calibration. The flux-calibrated spectra in the blue and the red regions were combined after scaling to get the final spectrum on a relative flux scale. The spectroscopic fluxes were checked with the photometric measurements at similar epochs and  appropriate scaling factors were applied. 
However, because of the location of the SN projected over the bright galaxy nucleus the background subtraction was a major problem. During the SN brightest stages, a careful reduction and choice of the background produced satisfactory results (still leaving a residual contamination, cfr. Sect.~\ref{sec:ha_profile}). At late stages (after about 200 d) the residual background contamination cannot be neglected. The observed continuum and the unresolved nebular emission lines are likely due to the host galaxy nucleus, while only the broad emissions, in particular H$\alpha$ and H$\beta$, are due to the SN.

\begin{table*}
\caption{Star ID and the magnitudes in {\it UBVRI} filters of the 9 secondary standards in the field of SN 2012ab}
\centering
\smallskip
\begin{tabular}{c c c c c c c c}
\hline \hline
Star ID  &$\alpha$      &$\delta$         	&  $U$    	    &  $B$              &  $V$          &  $R$          &  $I$                       \\
         & (h:m:s)      &($^{\circ}$ ' '')	&  (mag)  	    & (mag)             & (mag)         & (mag)         & (mag)                     \\
\hline                           
1	& 12:22:49.98	& +05:36:13.74	& 15.529  $\pm$  0.060	& 15.120 $\pm$ 0.030 & 14.321 $\pm$ 0.009  & 13.809 $\pm$  0.021 & 13.625  $\pm$ 0.038 	\\
2       & 12:22:52.79   & +05:34:36.87  & 18.316  $\pm$  0.066	& 18.033 $\pm$ 0.015 & 17.202 $\pm$ 0.025  & 16.686 $\pm$  0.055 & 16.299  $\pm$ 0.015 	\\
3       & 12:22:46.46   & +05:34:01.88  & 17.361  $\pm$  0.046	& 17.360 $\pm$ 0.015 & 16.666 $\pm$ 0.030  & 16.273 $\pm$  0.040 & 15.934  $\pm$ 0.011 	\\
4       & 12:22:35.55   & +05:34:55.02  & 20.276  $\pm$  0.194	& 19.297 $\pm$ 0.045 & 18.284 $\pm$ 0.034  & 17.645 $\pm$  0.075 & 17.140  $\pm$ 0.028 	\\
5       & 12:22:31.55   & +05:35:05.26  & 16.390  $\pm$  0.050	& 15.000 $\pm$ 0.020 & 13.634 $\pm$ 0.003  & 12.684 $\pm$  0.027 & 12.170  $\pm$ 0.241 	\\
6	& 12:22:35.89   & +05:37:34.51  & 15.604  $\pm$  0.043	& 15.926 $\pm$ 0.042 & 15.422 $\pm$ 0.031  & 15.082 $\pm$  0.002 & 14.825  $\pm$ 0.056 	\\
7	& 12:22:37.67   & +05:37:21.39  & 16.229  $\pm$  0.007	& 15.960 $\pm$ 0.016 & 15.133 $\pm$ 0.065  & 14.571 $\pm$  0.039 & 14.254  $\pm$ 0.026 	\\
8	& 12:22:44.94   & +05:38:11.86  & 18.911  $\pm$  0.073	& 17.802 $\pm$ 0.018 & 16.638 $\pm$ 0.016  & 15.907 $\pm$  0.061 & 15.427  $\pm$ 0.012 	\\
9	& 12:22:52.35   & +05:38:17.58  & 18.852  $\pm$  0.088	& 18.272 $\pm$ 0.020 & 17.350 $\pm$ 0.014  & 16.762 $\pm$  0.046 & 16.297  $\pm$ 0.016 	\\
\hline                                   
\end{tabular}
\label{tab:optical_observations}  
\newline
{The magnitudes quoted are the weighted average of the magnitudes obtained from Ekar 1.82m+AFOSC and HCT 2.0m+HFOSC}
\end{table*}                

\begin{table*}
\caption {Log of spectroscopic observation. The phase is measured after the explosion (MJD=55955.20$\pm$0.5); the resolution was
estimated by the FWHM of the night-sky lines [O I] at $\lambda$5577 \AA~or [O I] at $\lambda$6300 \AA. \label{tab:spec_obs}}
\begin{center}
\small\addtolength{\tabcolsep}{-2pt}
\begin{tabular}{c c c c c c}
\hline
{\bfseries Date} & {\bfseries Phase}  & {\bfseries MJD} &{\bfseries Telescope} & {\bfseries Range } & {\bfseries Resolution} \\
{\bfseries (dd/mm/yy)}&{\bfseries (d)}&  & & {\bfseries (\AA)} & {\bfseries (\AA)}\\
 \hline
\hline
07/2/12$^\dagger$ & 9.6 & 55964.85 & HET+LRS+gm300   & 4200-10200  &   10.9 \\
16/2/12$^\dagger$  & 18.6 & 55973.82 &  HET+LRS+gm300 &  4200-10200  &   10.9 \\
01/3/12$^\dagger$  &  32.6 & 55987.78 & HET+LRS+gm300 & 4200-10200 & 10.9 \\
07/3/12 & 38.3 & 55993.50  & Pennar+B\&C+300tr/mm & 3300-7800 & 7.5 \\
12/03/12& 43.4 & 55998.63 &Calar-Alto+CAFOS+gm8 & 3300-8850 & 13.0\\
14/03/12& 45.3  & 56000.54  &Copernico+AFOSC+gm4+gm2 & 3750-9900 &12.5-34.0 \\
15/03/12 & 46.5 & 56001.72 &NTT+SOFI+GB & 9350-24800 & - \\
17/03/12 & 48.5 & 56003.66 &Copernico+AFOSC+gm4& 3750-8200 & 11.5 \\
20/03/12 &  51.4     & 56006.63& NOT+ALFOSC+gm4 & 3450-9100 & 13.5 \\
24/03/12 &  56.3 &  56011.46 & Calar-Alto+CAFOS+gm8 & 3350-9900 & 14.0  \\
28/03/12 & 60.2 &  56015.41 &  Calar-Alto+CAFOS+gm8& 3350-10000 & 13.5  \\
28/03/12 & 60.2 & 56015.42 & Copernico+AFOSC+gm4 &  3400-8200 &    23.0 \\
14/04/12 & 76.5  & 56031.74 & NTT+EFOSC2+gm13 & 3650-9250 &  27.0 \\
25/04/12 & 86.3 & 56041.54 & Copernico+AFOSC+gm4 &   3600-8200  &  24.0 \\
30/04/12 & 93.3 & 56048.48 & NOT+ALFOSC+gm4 &  3450-8700 &   13.5 \\
09/05/12 &  102.2    & 56057.38 & Pennar+B\&C+300tr/mm & 3250-7750 &  7.5 \\
17/05/12 &  110.2  & 56065.40 & Pennar+B\&C+300tr/mm &   3300-7750 &   7.5 \\
26/05/12 & 119.2  &  56074.44 & WHT+ISIS+R300B+R158R & 3100-10550 & 4.3-7.3 \\
03/06/12 & 127.2  & 56082.42  & TNG+NICS+IJ+HK & 8750-24800  & - \\
08/06/12 & 132.2  &  56087.40 & TNG+LRS+LR-B  & 3150-8000  &  10.0\\
27/06/12 & 151.2  & 56106.39  & Copernico+AFOSC+gm4 & 3900-8150  & 24.0   \\
17/07/12 & 171.1  & 56126.35 & Copernico+AFOSC+gm4 &  3900-8150 & 24.0  \\
09/02/13 & 376.9  & 56332.09 & Copernico+AFOSC+gm4 &  3900-8150 & 24.0  \\               
16/03/13 & 411.9  & 56367.08 & Copernico+AFOSC+gm4 &  3900-8150 & 24.0  \\ 
25/04/14 & 817.3  & 56772.51  & Pennar+B\&C+300tr/mm &   3300-7750 &   7.5 \\     
27/02/15 & 1126.9 & 57080.15  & GTC+OSIRIS+R1000B    &   3630-7500 &   7.0 \\
\hline
\end{tabular}
\newline
$^\dagger$ Spectra from Biliniski et al. 2018
\end{center}
\end{table*}

\section{Temporal evolution of SN 2012ab}
\label{sec:lc}
\begin{figure*}
	\begin{center}
		\includegraphics[scale=0.50]{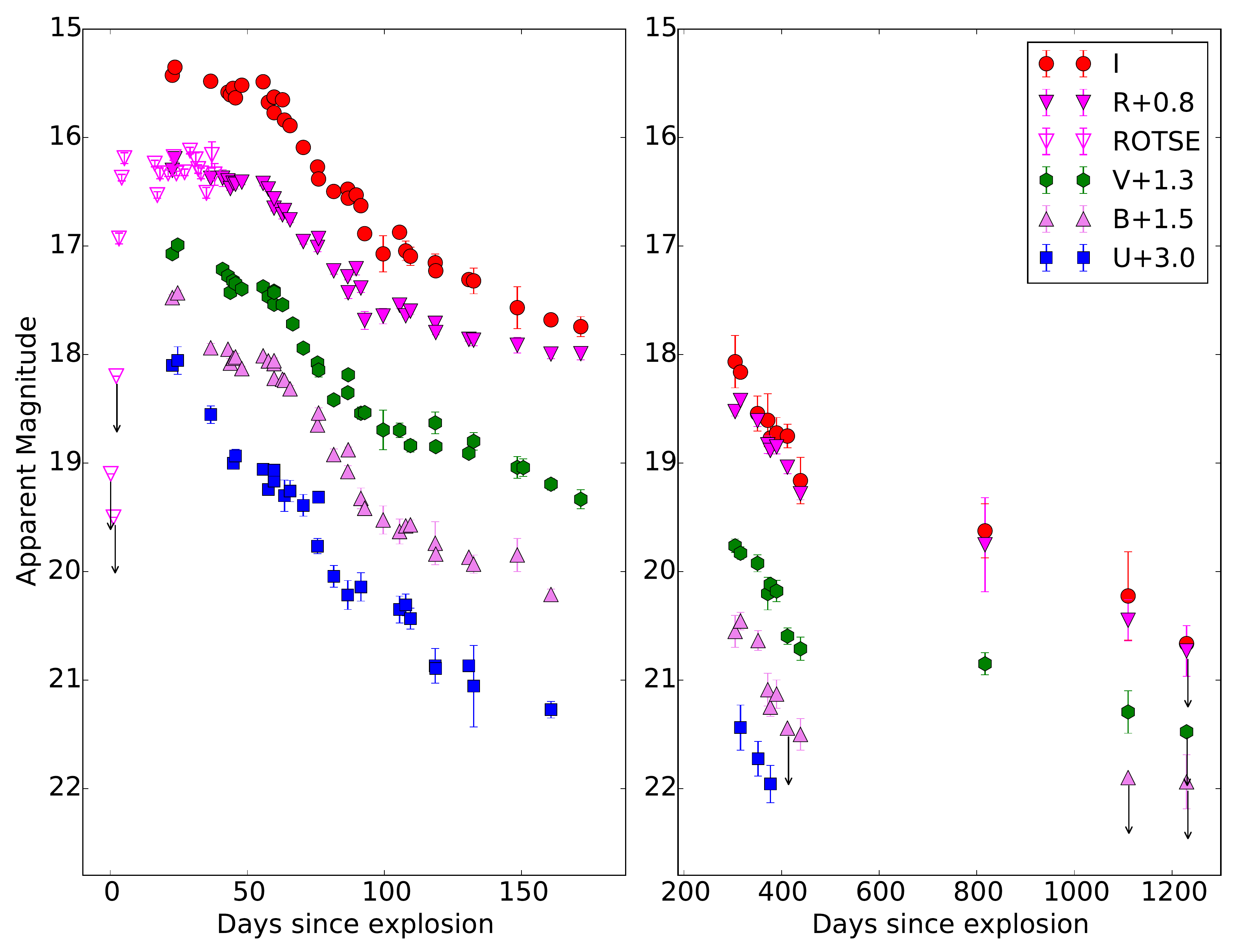}
	\end{center}
	\caption{Light curve evolution of SN 2012ab. The two panels show the  early (left) and late (right) evolution of SN 2012ab. The adopted reference date is that of explosion (MJD $=55955.3\pm0.5$; 2012 January 29). We have added the ROTSE-III \citep{2018MNRAS.475.1104B} data to our early $R$-band observations that are represented by empty symbols. The offsets in magnitudes applied to the {\it R, V, B} and $U$ data are 0.8, 1.3, 1.5 and 3.0 respectively.}
	\label{fig:lc}.
\end{figure*}

Optical light curves of SN 2012ab are shown in Fig. \ref{fig:lc}.  
This includes our new observations (Table \ref{tab:observation_log}),  that started on 2012 February 20 UT, and the ROTSE-IIIb observations (cfr. Sect.~\ref{sec:phot}) describing the early epochs. Thanks to the relatively deep and stringent pre-detection limits and to the ROTSE-IIIb detection on the rising branch, we could estimate the explosion date to be 2012 January $29.3\pm1$ (MJD 55955.3) which is used throughout the paper as reference date.
The light-curve in $R$-band shows a rapid rise of $\geq$ 3.4 mag in the first 4 d reaching a $R_{max}=15.39$ on MJD 55960.34. After the rise, as noted by \cite{2018MNRAS.475.1104B} the light curve remained almost flat for about 60 d. In other bands the coverage started only on February 20.
In {\it B, V, R} and $I$ bands we can identify a short plateau lasting about 60 d, steeper at shorter wavelengths (from 1.62 to 0.54 \mhund), followed by a rapid drop in the next 40 d (ranging from 3.67 to 2.72 \mhund).
The $U$ band instead declines at constant rate from the early available observations to 100 d (2.97 \mhund).
The slopes of the different phases identified in the light curves are reported in Table \ref{tab:comp_decay_rate}.  
In the 130$-$350 d interval, the decline at all wavelengths is definitely slower (from 0.17 to 0.42 \mhund) than the typical radioactive decay rate of the $^{56}$Co $\rightarrow$ $^{56}$Fe (0.98 \mhund) indicating the contribution  of another energy source in addition to radioactive decay, most probably the interaction between the SN ejecta and the CSM, as discussed in the following Sections. 
Only for a short interval between 350$-$440 d, all light curves decline at a rate close to that of $^{56}$Co (between 0.62 and 0.93 \mhund), to slow-down once again afterward in {\it V, R} and {$I$}, the bands for which we have late time observations.

\begin{table*}
\caption{Decay rates of the light curves of SN 2012ab in \mhund.}
\centering
\smallskip
\begin{tabular}{c c c c c c c c c c}
\hline \hline
Time Interval & 20$-$60d & 20$-$100d & 60$-$100d & 130$-$350d & 350$-$440d &  440$-$1110d\\
    &                  &             &           &            &            &             \\
\hline
$U$ & ---             & 2.97 $\pm$ 0.20 & ---         & 0.17 $\pm$ 0.09 & 0.92 $\pm$ 0.00   & --- \\ 
$B$ & 1.62 $\pm$ 0.27 & ---         & 3.67 $\pm$ 0.23 & 0.33 $\pm$ 0.07 & 0.89 $\pm$ 0.03   & --- \\
$V$ & 1.18 $\pm$ 0.13 & ---         & 2.95 $\pm$ 0.30 & 0.42 $\pm$ 0.04 & 0.93 $\pm$ 0.13  & 0.08 $\pm$ 0.02  \\
$R$ & 0.59 $\pm$ 0.17 & ---         & 2.72 $\pm$ 0.16 & 0.34 $\pm$ 0.03 & 0.71 $\pm$ 0.08 & 0.17 $\pm$ 0.03 \\
$I$ & 0.54 $\pm$ 0.23 & ---         & 3.29 $\pm$ 0.17 & 0.31 $\pm$ 0.03 & 0.62 $\pm$ 0.15 & 0.16 $\pm$ 0.02 \\ 
\hline
\end{tabular}
\newline
\label{tab:comp_decay_rate} 
\end{table*}

\subsection{Comparison sample}
\label{sec:comp_sample}
To place SN 2012ab in the context of other SNe, we have selected a reference sample of other well-studied CCSNe. The sample is quite heterogeneous and comprises of nine objects: SNe 1996al, 1998S, 1999em 2005ip, 2006gy, 2007od, 2009kn,  2010jl and PS1-10adi. The basic parameters of the comparison SNe are tabulated in Table~\ref{tab:photometric_parameters_different_SNe}. SNe 2006gy and PS1-10adi  have been selected because they are located very close to the nuclei of their host galaxies, similar to SN 2012ab. This is motivated by the need of checking whether positional dependency can affect photometric or spectroscopic features. Then we have selected a group of strongly interacting SNe IIn similar to SNe 2005ip and  2010jl, whose light curve features are very similar to those of SN 2012ab. In contrast, we choose the fleeting SN~1998S, the interacting linear SN 1996al, the transitional (from Plateau to Linear) SN~2007od and the Type IIn-P like SN~2009kn, to remark the clear photometric differences with SN~2012ab for objects showing a rapid luminosity decline. At last, we have included a normal Type II Plateau, SN~1999em, to have a clear comparison with a non-interacting SN. Thus, the diversity in the comparison sample enables us to highlight the light curve heterogeneity of SNe II, and the role of CSM interaction in their photometric evolution.

\begin{table*}
\caption{Properties of the SNe IIn comparison sample }
\centering
\smallskip
\begin{tabular}{l c c c c c c}
\hline \hline
SN       & Host Galaxy      & Offset from center & Distance$^\ddagger$      & Extinction       & Peak M$_R$        & Reference$^\dagger$  \\
        &                  &                    & (Mpc)         &$E(B-V)$ (mag)    &  (mag)            &     \\
\hline
SN 1996al        &  NGC 7689        & 30" N              & 22.9            & 0.11       & $-$17.53            & 1 \\
SN 1998S        & NGC 3877         & 16" W, 46" S        & 15.7            & 0.219      & $-$19.54            & 2,3,4 \\
SN 1999em       & NGC 1637         & 15" W, 17" S            & 9.8              & 0.1  &  $-$16.93               & 5     \\
SN 2005ip$^{*}$ & NGC 2906         & 2".8 E, 14".2 N     & 29.3    & 0.048      & $-$17.44            & 6  \\
SN 2006gy       & NGC 1260         & 0.941" W, 0.363" N               & 79.7            & 0.56       & $-$22.39            & 7,8 \\
SN 2007od       & UGC 12846        & 38" E, 31" S        & 25.6    & 0.126      & $-$18.61            & 9,10   \\   
SN 2009kn       & MCG -3-21-6      & 17.75" E, 15.27" N                & 67.5            & 0.114      & $-$18.09            & 11   \\
SN 2010jl$^{*}$ & UGC 5189A       & 2".4 E, 7".7 N      & 48.5    & 0.058      & $-$20.25            &  12, 13,14    \\
PS1-10adi$^{*}$ & J204244.74+153032.1&      center      & 953.4           & 0.091      & $-$23.16            & 15 \\
\hline                                                                              
\end{tabular}
\newline
$^\dagger$ References:(1) \cite{2016MNRAS.456.3296B};  (2) \cite{2000MNRAS.318.1093F}; (3) \cite{2016AJ....152...50T}; (4) \cite{2001MNRAS.325..907F}; (5) \cite{2003MNRAS.338..939E}; (6) \cite{2012ApJ...756..173S}; (7) \cite{2007ApJ...666.1116S}; (8) \cite{2009ApJ...691.1348A}; (9) \cite{2011MNRAS.417..261I};  (10)   \cite{2010ApJ...715..541A}; (11) \cite{2012MNRAS.424..855K};  (12) \cite{2016MNRAS.456.2622J}; (13) NED; (14) \cite{2012AJ....144..131Z}; (15) \cite{2017NatAs...1..865K};

$*$ We use days since discovery with the assumption that the discovery date occurred close to the explosion date.\\
$^\ddagger$ Distances have been scaled to H$_0$ = 73 km sec$^{-1}$ Mpc$^{-1}$
\label{tab:photometric_parameters_different_SNe}      
\end{table*}

\subsection{Absolute magnitude and bolometric light curve}
\label{sec:abs_bol}

\begin{figure}
	\begin{center}
		\includegraphics[width=1.0\linewidth]{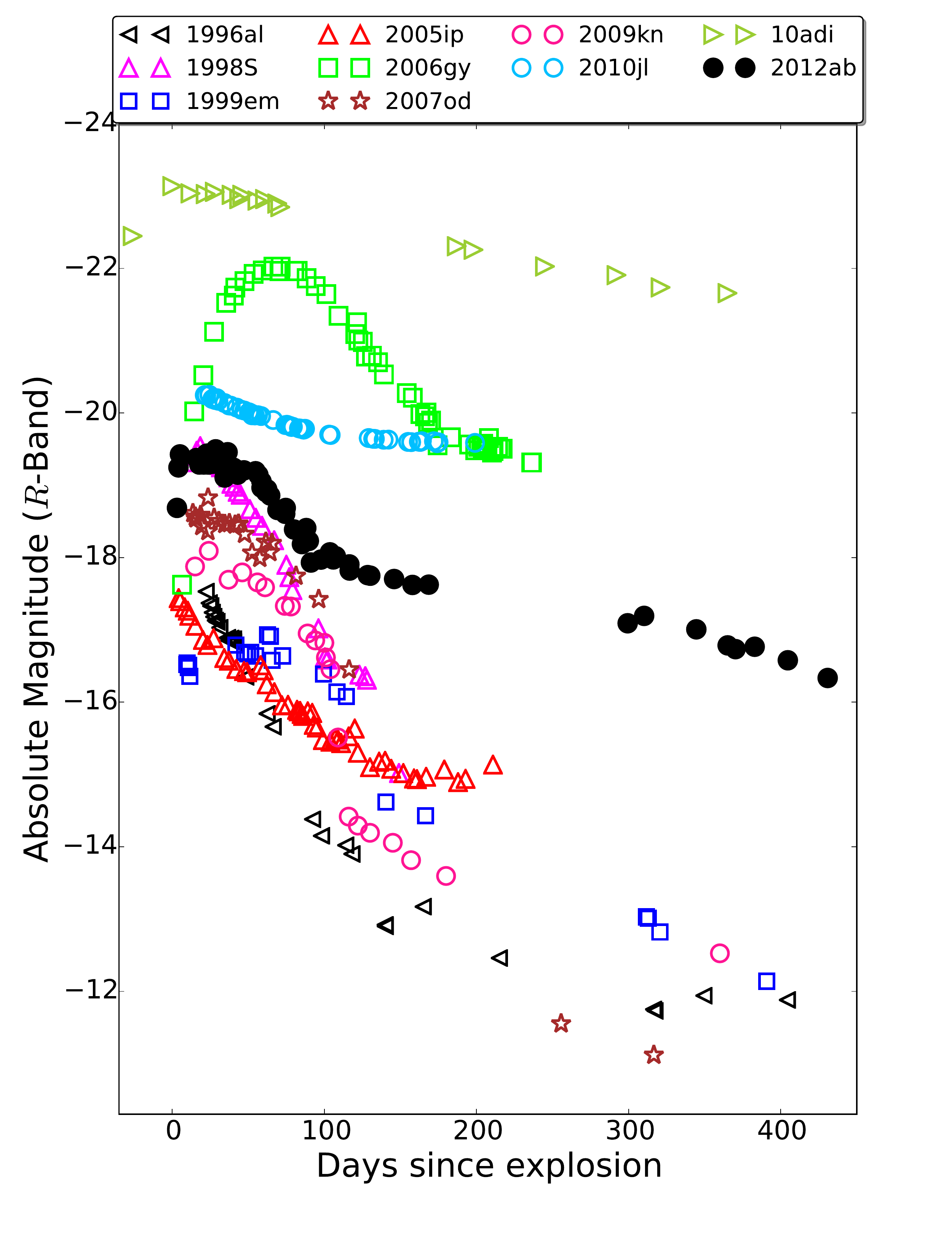}
	\end{center}
	\caption{Absolute magnitude ($R$-band) light curve of SN 2012ab as compared with those of the reference sample.  Time dilation correction has been applied to the light curve of SN 2012ab. }
	\label{fig:absmag}
\end{figure}

Adopting the distance modulus $\mu=34.57$ mag and a total reddening $E(B-V)=0.079$ mag, we can calculate the peak absolute magnitude. The absolute magnitude in the $R$ band is M$_R= -19.39$ mag, computed by using the early brightest point from ROTSE-IIIb (MJD 55960.34;  R=15.39). Fig.~\ref{fig:absmag} shows the comparison of the absolute R-band light curve of SN 2012ab with those of the SNe of the comparison sample. SN 2012ab has a peak luminosity similar to SN 1998S (cfr. Table~\ref{tab:photometric_parameters_different_SNe}).

To better understand the overall photometric properties of SN 2012ab and - in particular - the sources of energy at work, we computed the pseudo-bolometric (limited to BVRI bands) light curve, thanks to the multi-wavelength follow-up described in the previous Sections. Broad-band magnitudes were converted into fluxes at the effective wavelengths, corrected for extinction, and finally the resulting Spectral Energy Distribution (SED) was integrated over wavelengths. The fluxes were computed at the epochs when R-band observations were available. When an observation in one band was not available on a given night, the magnitude has been obtained by interpolating the light curve with a low-order polynomial, or was extrapolated assuming a constant colour.  
The pseudo-bolometric light curve of SN 2012ab shows the same behaviour of the chromatic light curves, i.e. the rapid rise to maximum, the plateau up to about 2 months and the steep fall from the plateau. At about 4 months the decline slows down and remains slower than the radioactive decay of $^{56}$Co up to the latest detection (3.3 yr after explosion). The fastest late-time decline occurs between 350$-$440 d with a slope of $0.78\pm0.08$ \mhund. It is clear that another energy input is required in addition to the $^{56}$Co radioactive decay to sustain the light curve. We will see in Section 5 that the ejecta-CSM interaction provides a plausible explanation.
Assuming that the luminosity of SN 2012ab in the 350$-$440 d period was generated by the radioactive decay of $^{56}$Co, the $^{56}$Ni mass required is M($^{56}$Ni)$\geq0.97$ \Msun, using the formulation given by \citet{2003ApJ...582..905H}. This value, much larger than the usual $^{56}$Ni masses produced in normal CCSNe \citep[0.001-0.03 \Msun][]{2003ApJ...582..905H}, is also excluded by a simple model based on \citet{1982ApJ...253..785A} formalism. In fact, if we assume that the luminosity of the SN in the 350$-$440 d is entirely supported by the $^{56}$Ni decay, then following \citet{1982ApJ...253..785A}, the early rise is matched by only assuming 1~\Msun~ of ejecta, e.g. an almost pure $^{56}$Ni composition, which is not supported by the observations. Higher ejecta mass would give even slower rise times. Also the rapid decline seen in the bolometric light curve after 100d is at odds with a high $^{56}$Ni mass scenario.

\begin{figure}
	\begin{center}
		\includegraphics[width=1.0\linewidth]{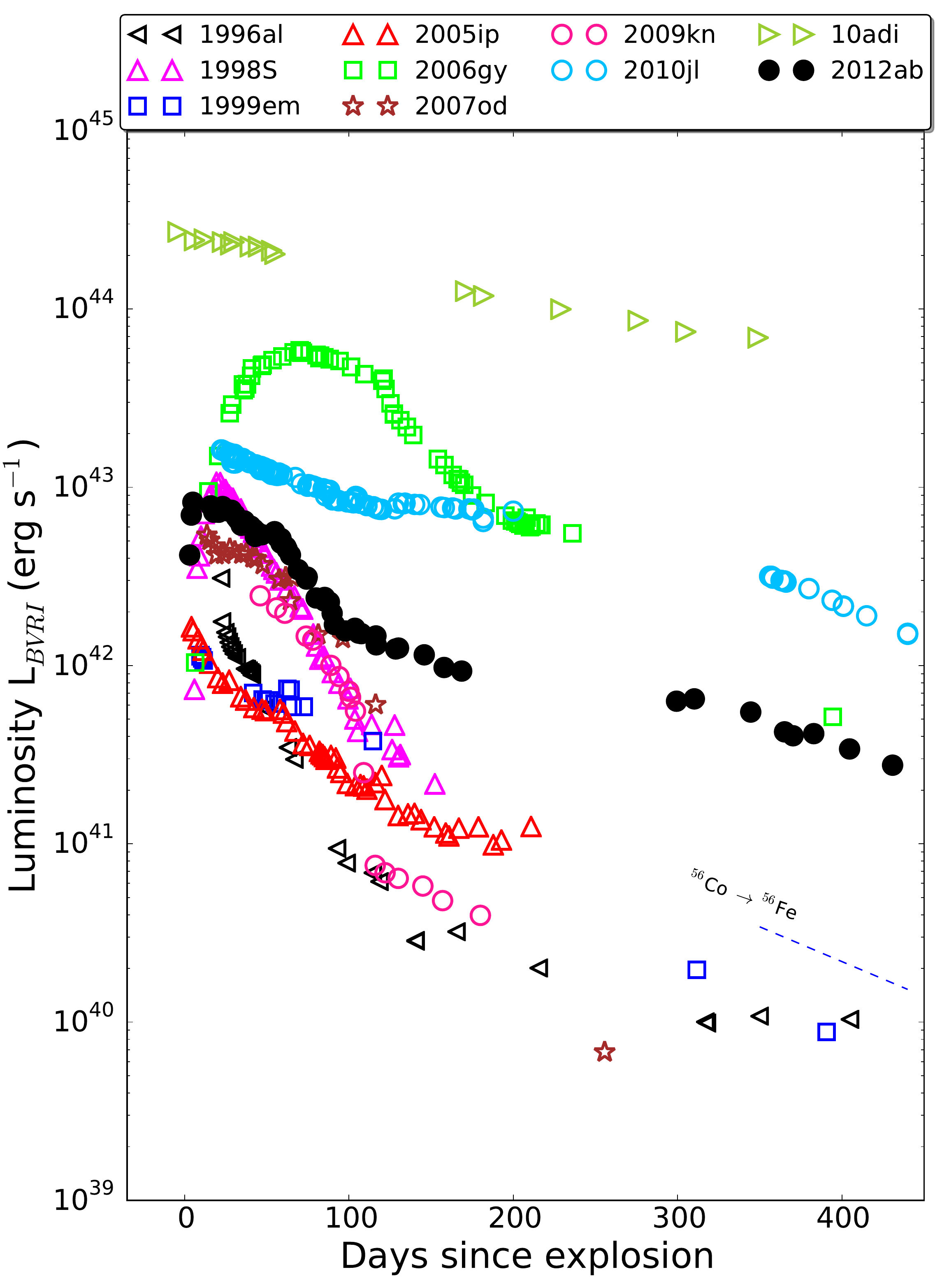}
	\end{center}
	\caption{Pseudo-bolometric ({\it BVRI}) light curve evolution of SN 2012ab as compared with those of the reference sample. The dashed line shows the expected decline rate of $^{56}$Co to $^{56}$Fe. Time dilation correction has been applied to the light curve of SN 2012ab.}
	\label{fig:bol}
\end{figure}

In Fig.~\ref{fig:bol} the pseudo-bolometric light curve of SN 2012ab is compared with those of the comparison sample.
SN 2012ab has average peak luminosity with respect to the comparison objects (L$_{BVRI}^{max}\simeq8.24\times10^{42}$ erg s$^{-1}$). It is fainter than SNe 2006gy, 2010jl and PS1-10adi, brighter than SNe 1999em, 2005ip and 2009kn, and similar to SNe 1996al, 1998S and 2007od. However, among these last four objects the luminosity evolution varies significantly and the behaviour most similar to SN 2012ab is with that of the interacting SNe group. 

\subsection{Colour evolution of SN 2012ab}
\label{sec:colour}
\begin{figure}
	\includegraphics[width=\columnwidth]{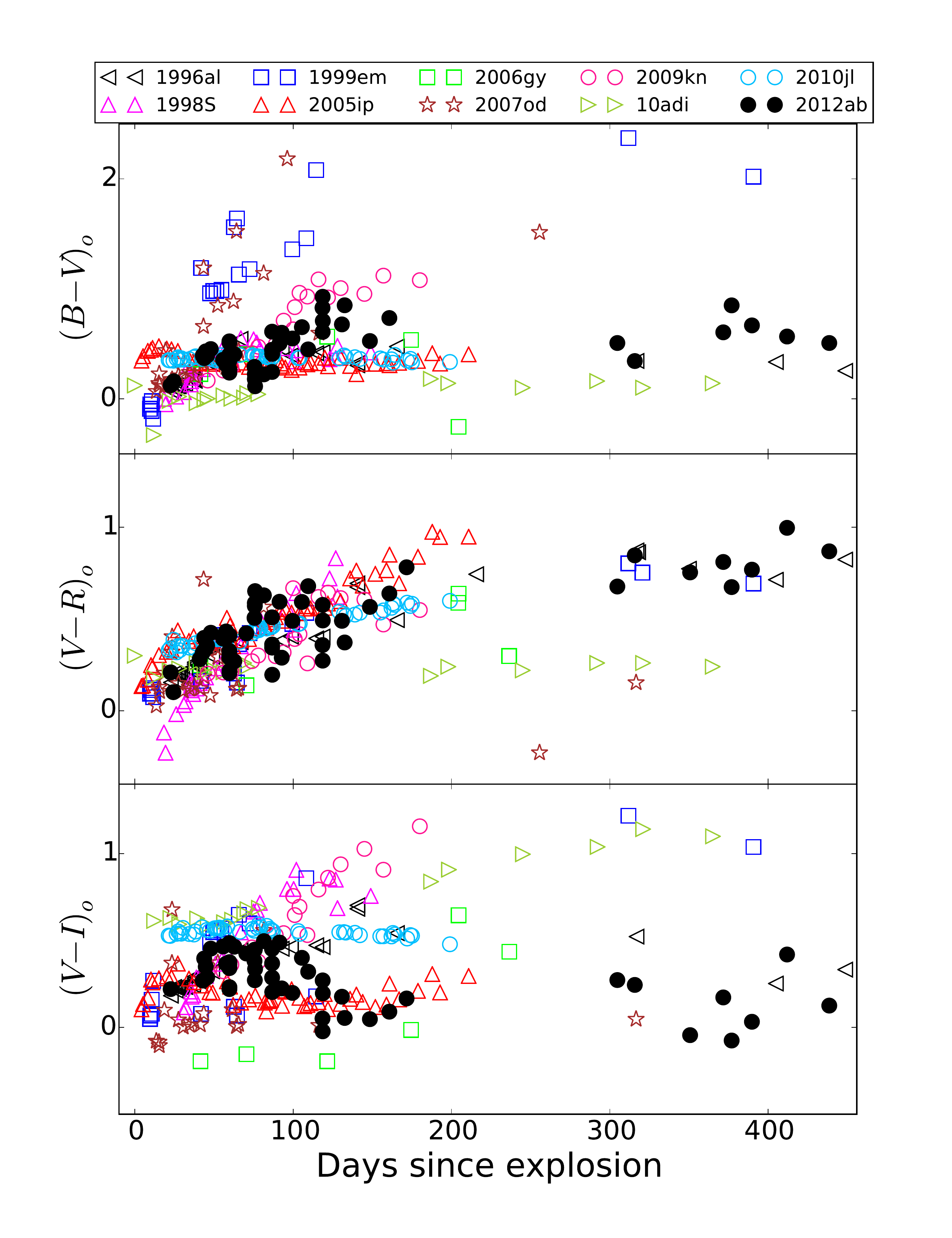}
	\caption{Reddening-corrected colour curve evolution of SN 2012ab as compared with those of the reference sample.}
	\label{fig:colorcurve}
\end{figure}

The reddening-corrected colour curves of SN 2012ab are compared  in Fig. \ref{fig:colorcurve} with those of the reference sample.

The {\it (B$-$V)} colour  of SN 2012ab evolves redward very slowly, from $\sim$ 0.1 mag to $\sim$ 0.7 mag between 22 and 120 d. This reflects in a slow reddening of the spectral continuum during this period (cfr. Sect.~\ref{sec:rad_temp}). 
A similar, quite constant trend in {\it (B$-$V)} colour evolution is also noticed for the other strongly interacting SNe 1996al, 1998S 2005ip, 2006gy, 2010jl and PS1-10adi \citep{2009ApJ...695.1334S,2016MNRAS.456.2622J}, indicating that the colour evolution is not driven by the expansion of the ejecta as is in the cases of the normal SN 1999em and the weakly interacting SN 2007od that show a fast drift to the red. SN 2009kn appears intermediate between normal and strongly interacting SNe. 
After 120 d, up to the latest epoch (440 d), the {\it (B$-$V)} colour of SN 2012ab does not change significantly, with a possible shallow decrease ($(B-V)\sim 0.35$ mag) at intermediate epochs, in line with other interacting SNe, while normal SN II remain definitely redder.

Less evident, though still present, is the differentiation of the {\it (V$-$R)} colour between normal and interacting objects. 
SN 2012ab changes slowly, similarly to interacting SNe. The {\it (V$-$R)} colour of SN 2012ab becomes progressively redder, reaching $(V-R)\sim 0.9$ mag at 440 d, consistent with the strengthening of H$\alpha$ line relative to the continuum at late times (see Section \ref{sec:spec}). 
Two very rare additional points (not shown in Fig. \ref{fig:colorcurve}) indicate a return to a bluer {\it (V$-$R)} colour up to 1100 d.
Different is the {\it (V$-$I)} behaviour of SN 2012ab, which remains relatively almost constant colours throughout the whole evolution with a modest drift to the red around 100d, and then fixing  at $(V-I)\sim 0.05$ mag up to 440 d. The latest points (not shown in Fig. \ref{fig:colorcurve}) seems to indicate much bluer {\it (V$-$I)} colours but at epoch when the photometric errors are large.

\section{Spectral evolution of SN 2012ab} 
\label{sec:spec}
\begin{figure*}
		\includegraphics[width=1.0\textwidth]{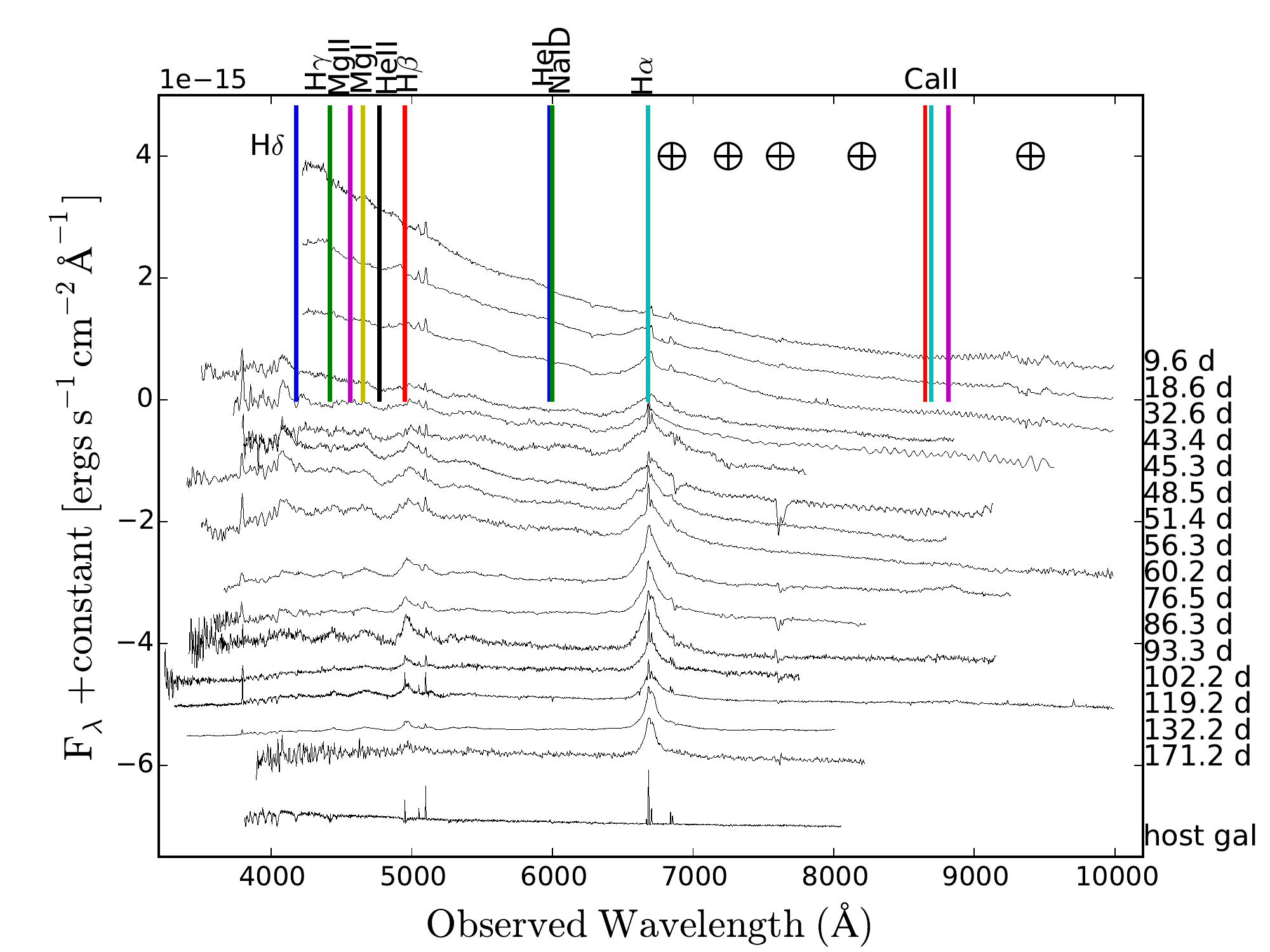}
	\caption{Spectral evolution of SN 2012ab from 9.6 d to 171.2 d. The first 3 spectra are from \citet{2018MNRAS.475.1104B}. All spectra are displayed in the observer restframe and are not corrected for reddening. 
	The flux scale is relative to the first spectrum while other spectra have been downshifted by $0.41\times10^{-15}$ erg s$^{-1}$ cm$^{-2}$ A$^{-1}$ for better display. The regions of the main telluric absorption features are also marked on the top.}
	\label{fig:earlyspec}
\end{figure*}

\begin{figure}
            \includegraphics[width=1.0\linewidth]{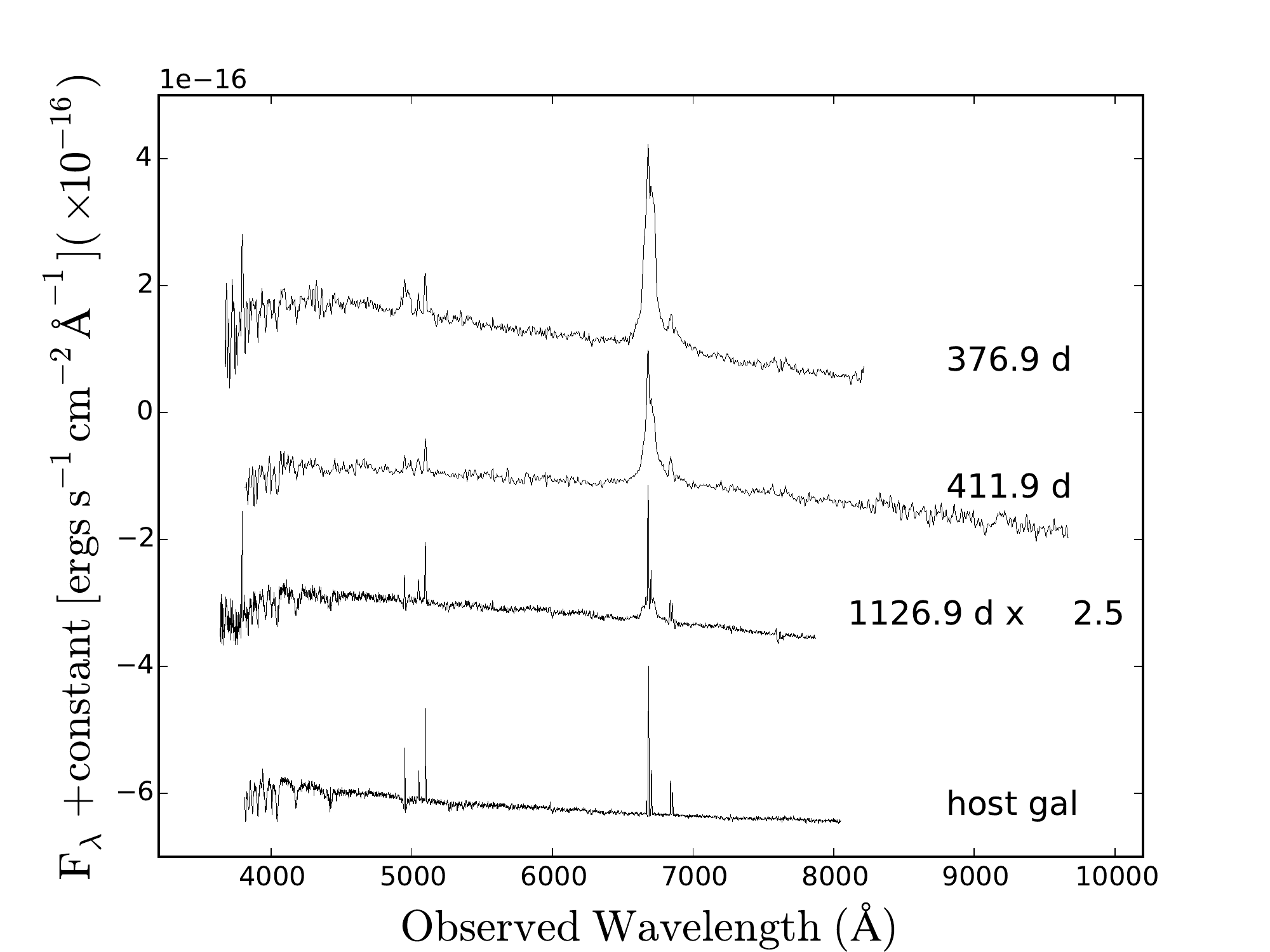}
	\caption{Late phase spectral evolution of SN 2012ab.
	All spectra are displayed in the observer restframe and are not corrected for reddening. 
	The flux scale is relative to the first spectrum. For better display other spectra have been downshifted by $0.2\times10^{-15}$ erg s$^{-1}$ cm$^{-2}$ A${^{-1}}$ and the spectrum on 1126 d multiplied by 2.5.}
	\label{fig:latespec}
\end{figure}

The spectral evolution of SN 2012ab is shown in  Figs. \ref{fig:earlyspec} and \ref{fig:latespec}.

The main characteristic of SN 2012ab is its slow spectroscopic evolution after the rapid, initial cooling of the continuum.
The spectra are dominated at all epochs by emission lines, in particular \Ha, whose complex evolution will be discussed in Section \ref{sec:ha_profile}.
Only a broad and shallow P-Cygni absorption leftwards of H$\alpha$ and H$\beta$ is seen in the first months.
The lines of Fe II, Ba II, Sc II, Mg I, Ti II, and Ca II, with P-Cygni profile, characteristic features of SNe are not visible at any time.
Narrow lines due to [O II], [O III], [N II] and [S II] are always well visible suggesting heavy host galaxy contamination.

The first spectrum taken 9.6 d after the explosion is an almost featureless black-body continuum at 12600 K with  the above mentioned unresolved emission lines of the galaxy nucleus and, possibly, a shallow \Ha\/ absorption (cfr. Sect.~\ref{sec:ha_profile} and Table~\ref{tab:gauss_log}).
With time the temperature rapidly decreases.
The black-body temperature cools down to 10100$\degree$K at 18.6 d and broad asymmetric emissions of \Ha\/ and \Hb\/ emerge. 
By 32 d the shallow H$\alpha$ absorption becomes clearly visible (cfr. Sect.~\ref{sec:ha_profile}).

Afterward the evolution slows down significantly. The temperature slowly decreases (cfr. Sect.~\ref{sec:rad_temp}), a blend of emissions, possibly multiplet 42 of Fe II, develops redward of \Hb, the line width overall shrinks and the SN continuum fades.
The spectra at one year, and later, display only broad \Ha\/ and \Hb\/ emissions of the SN contaminated by narrow host galaxy features over a residual galaxy nucleus spectrum (cfr. Fig.~\ref{fig:latespec}). 
The nebular emission lines of intermediate atomic mass elements typical of late-time spectra of CCSNe (e.g. [O I], [Ca II], [O II]) are not present at any time.

\subsection{Infrared spectrum}
\label{sec:IR}
\begin{figure}
	\begin{center}
		\hspace{-1.0cm}
		\includegraphics[width=1.0\linewidth]{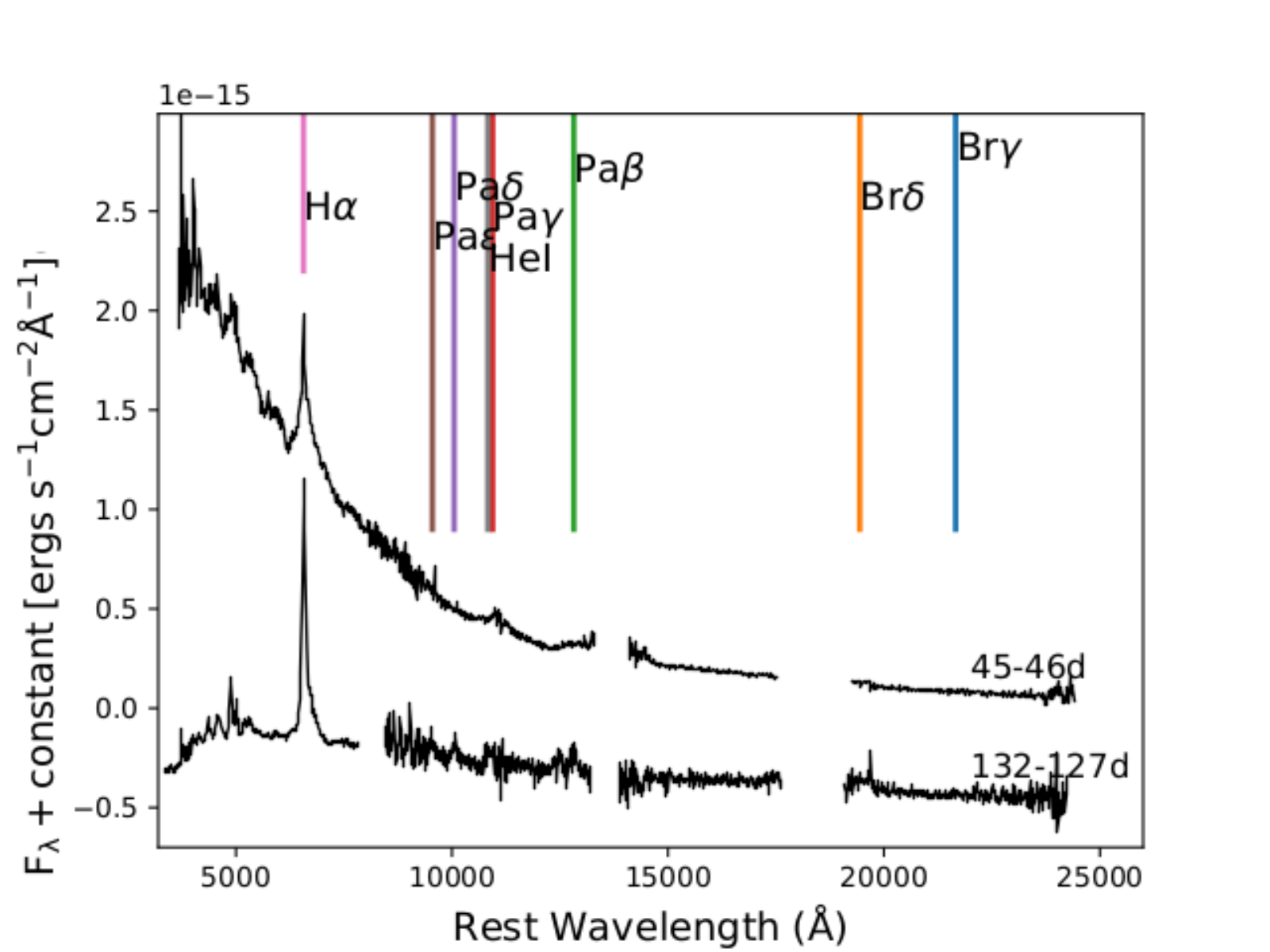}
	\end{center}
	\caption{The optical-NIR spectra of SN 2012ab at two phases. The epochs of the optical and NIR spectra are indicated, respectively. Spectra have been reported to the galaxy restframe and reddening corrected. For display the second spectrum has been down-shifted by $0.5\times 10^{-15}$ erg s$^{-1}$ cm$^{-2}$ A$^{-1}$. Main line identifications are reported.}
	\label{fig:NIR}
\end{figure}

Two NIR spectra have been acquired. They are displayed in Fig \ref{fig:NIR} along with the closest optical spectrum available.
The first spectrum obtained with closest available optical NTT+SOFI at 46 d from the explosion shows clearly two broad Paschen line, P$_{\beta}$ and P$_{\gamma}$, the latter probably blended with He I line at 10830 \AA. The FWHM width of the two Paschen lines (FWHM $\sim 17000-17500$ \kms) are similar to that of the central (IW) emission of \Ha, that will be discussed in  Sect.~\ref{sec:ha_profile}.
In the TNG+NICS spectrum (phase 127 d) four Paschen lines are visible (P$_{\beta}$, P$_{\gamma}$, P$_{\delta}$ and P$_{\epsilon}$). Again P$_{\gamma}$ is blended with He I at 10830 \AA and P$_{\epsilon}$. The other three lines are isolated and their FWHMs and positions are in perfect agreement with the values of the central (IW) emission of \Ha\/ (cfr. Sect.~\ref{sec:ha_profile} and Table~\ref{tab:gauss_log}).

\subsection{Spectral comparison with other SNe IIn}
\label{sec:comp_spec}

\begin{figure}
	\begin{center}
		\includegraphics[width=1.0\linewidth]{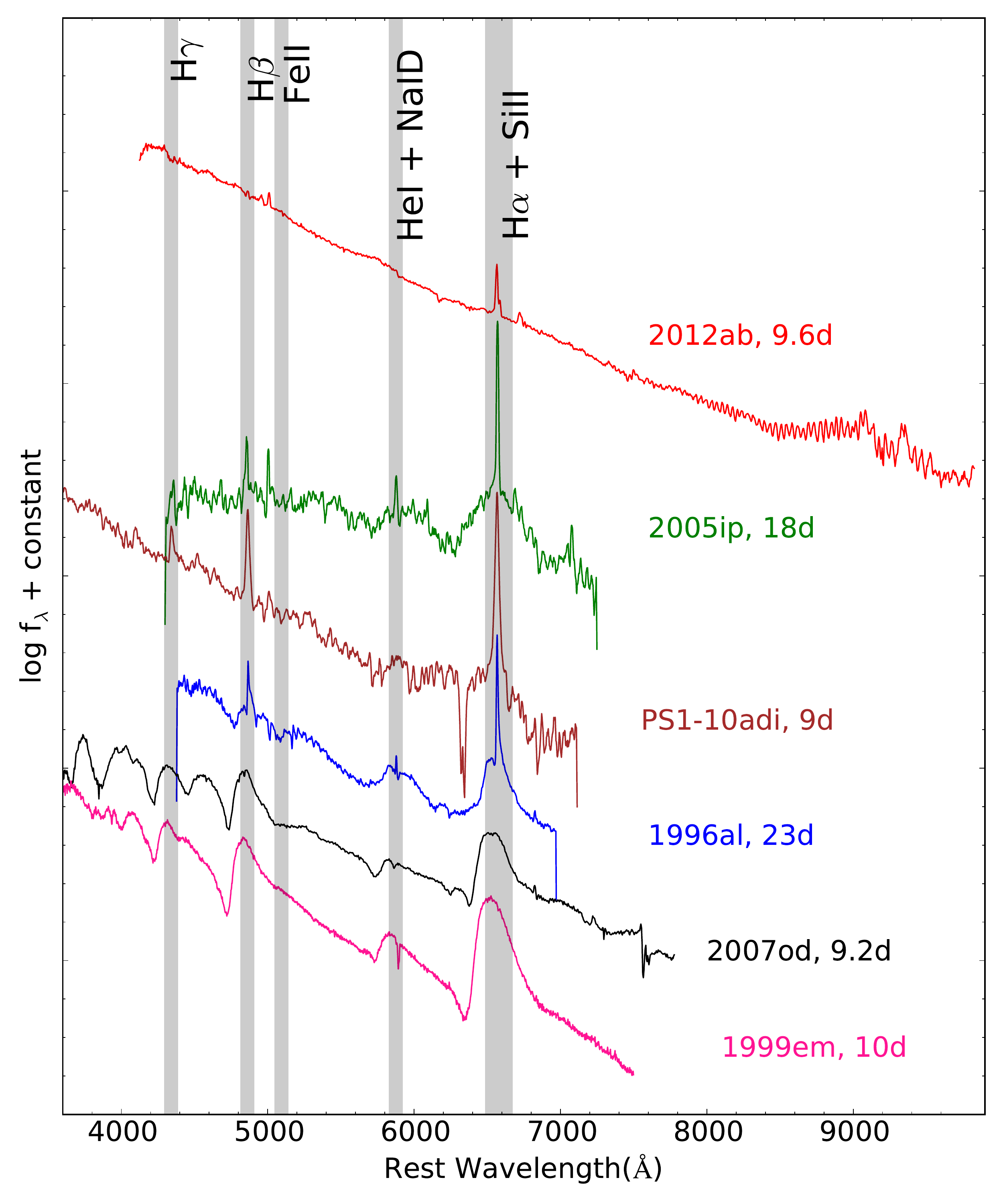}
	\end{center}
	\caption{Early time (about 9 d after the explosion) spectral comparison between SN 2012ab and the comparison sample. All spectra have been corrected for their host-galaxy recession velocities and for extinction, with the values reported in Table~\ref{tab:photometric_parameters_different_SNe}.}
	\label{fig:compearlyspec}
\end{figure}

\begin{figure}
	\begin{center}
		\includegraphics[width=1.0\linewidth]{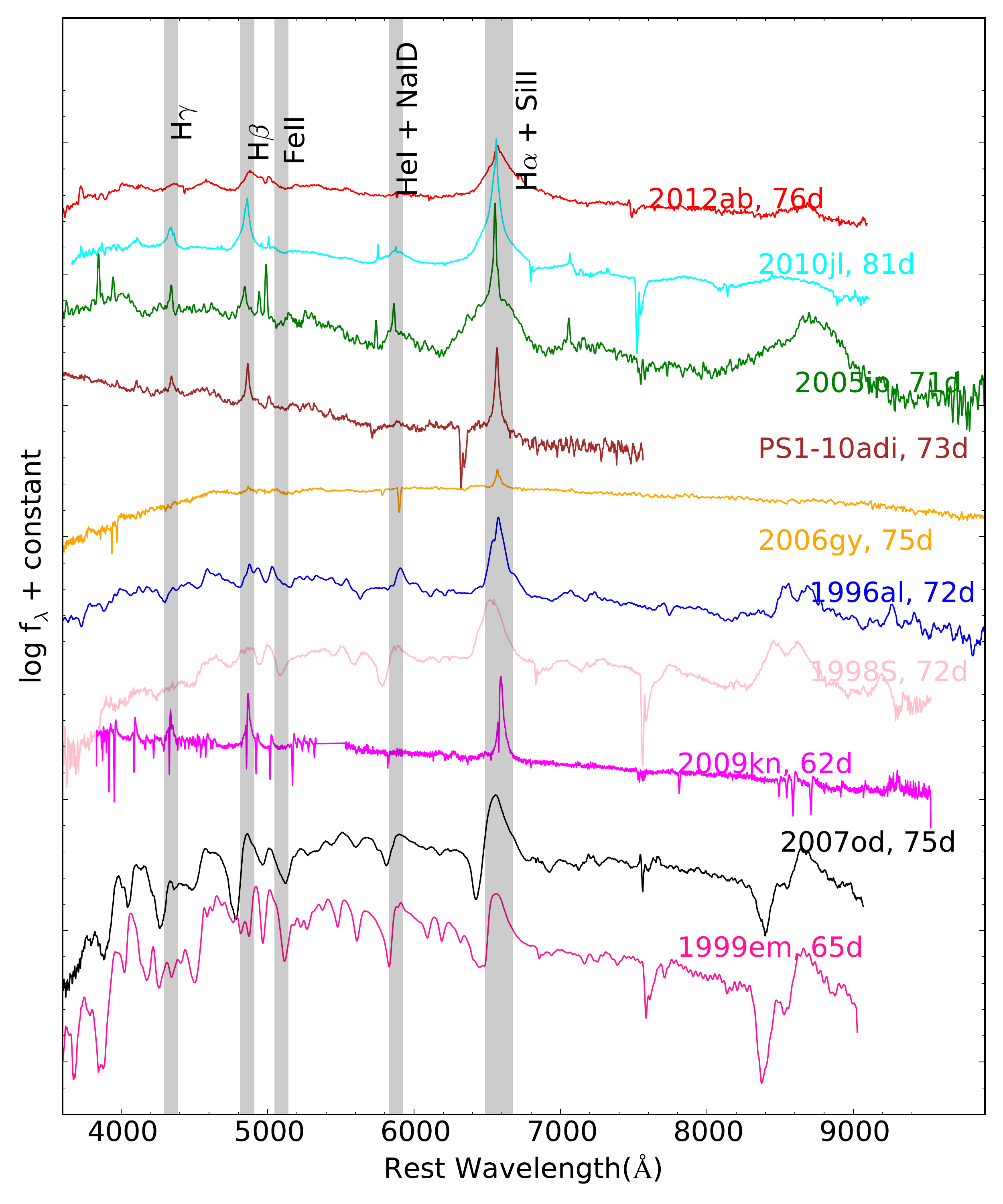}
	\end{center}
	\caption{Spectral comparison at about 76 d after the explosion between  SN 2012ab and the comparison sample. All spectra have been corrected for their host-galaxy recession velocities and for extinction.}
	\label{fig:compmidspec}
\end{figure}

\begin{figure}
	\begin{center}
		\includegraphics[width=1.0\linewidth]{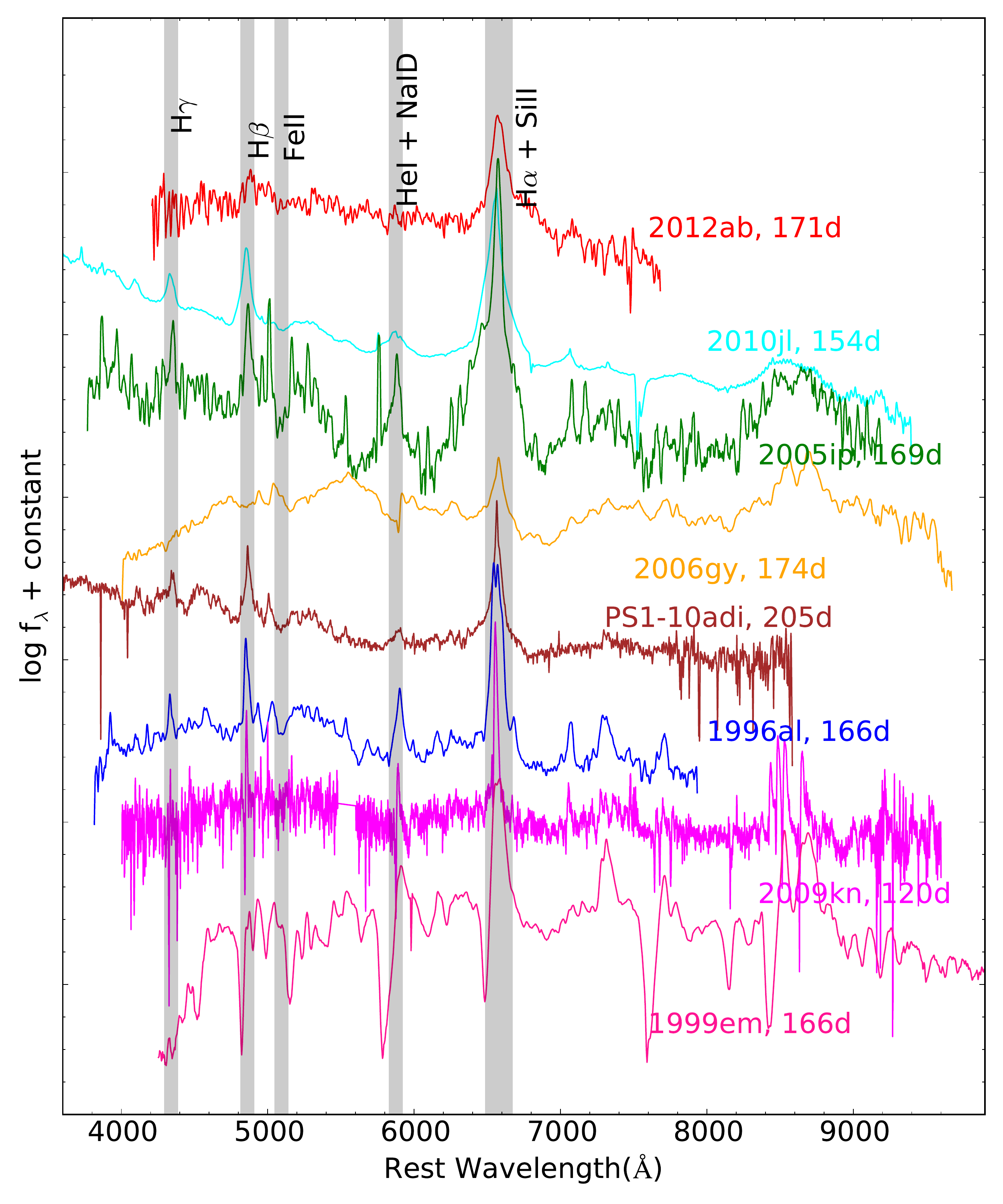}
	\end{center}
	\caption{Late-time (about 170 d after explosion) spectral comparison between SN 2012ab and the other objects of the comparison sample. All spectra have been corrected for their host-galaxy recession velocities and for extinction.}
	\label{fig:complatespec}
\end{figure}

In Figs.~\ref{fig:compearlyspec}, \ref{fig:compmidspec} and \ref{fig:complatespec} we compare the early, mid and late phase spectra of SN 2012ab with those of the reference sample of SNe IIn (see Section \ref{sec:comp_sample}). The plots are made with spectra taken at similar phases with respect to the explosion. For SN 2012ab, we plot the spectra without the additional removal of the galaxy contamination discussed in Sect. \ref{sec:ha_profile}. 

At early (9 d) phases all spectra show blue continua. However, Fig.  \ref{fig:compearlyspec} shows that SN 2012ab, almost featureless has low-contrast lines over the continuum, is similar to PS1-10adi. Some objects, such as SNe 2007od and 1996al, show more strong features shaped lines not dissimilar to normal Type II (Plateau or Linear) SNe, while others already display multicomponent \Ha\/ profiles (e.g. SN2005ip, though its phase is considerably later).
 
Also the mid-evolution (76 d) spectrum (Fig. \ref{fig:compmidspec}) shows features similar to the other strongly interacting SNe. In particular, evident is the complex profile of \Ha\/ emission, in analogy with SNe 2005ip, 2010jl, and PS1-10adi,  each exhibiting large expansion velocities.
Interestingly, SN~2012ab does not show the He I emission, a feature missing only in SN 2009kn.

About six months after the explosion SN 2012ab keeps an overall similarity with strongly interacting SNe (Fig \ref{fig:complatespec}). While the  interacting SN 1996al shows an unusual, strong He I emission and a three-peaked \Ha\/ emission, the strongly interacting ones have broader components, clearly visible in SNe 2005ip and 2010jl. The component deconvolution of SN~2012ab will be discussed in Sect. \ref{sec:ha_profile}. 
The strong interaction between SN ejecta and CSM produces also a blue pseudo-continuum at $\lambda$ $\leq$ 5500 \AA, that is likely produced by a forest of Fe emission lines \citep{2009ApJ...691..650F}. Such enhanced blue continuum, evident in SNe 2005ip, 2010jl and PS1-10ad, is not visible in SN 2012ab.

\subsection{Photospheric radius and temperature}  
\label{sec:rad_temp}

The analysis of the spectra allows to derive a number of physical parameters such as photospheric radius and temperature. 
 
\begin{figure}
	\begin{center}
		\includegraphics[width=1.0\linewidth]{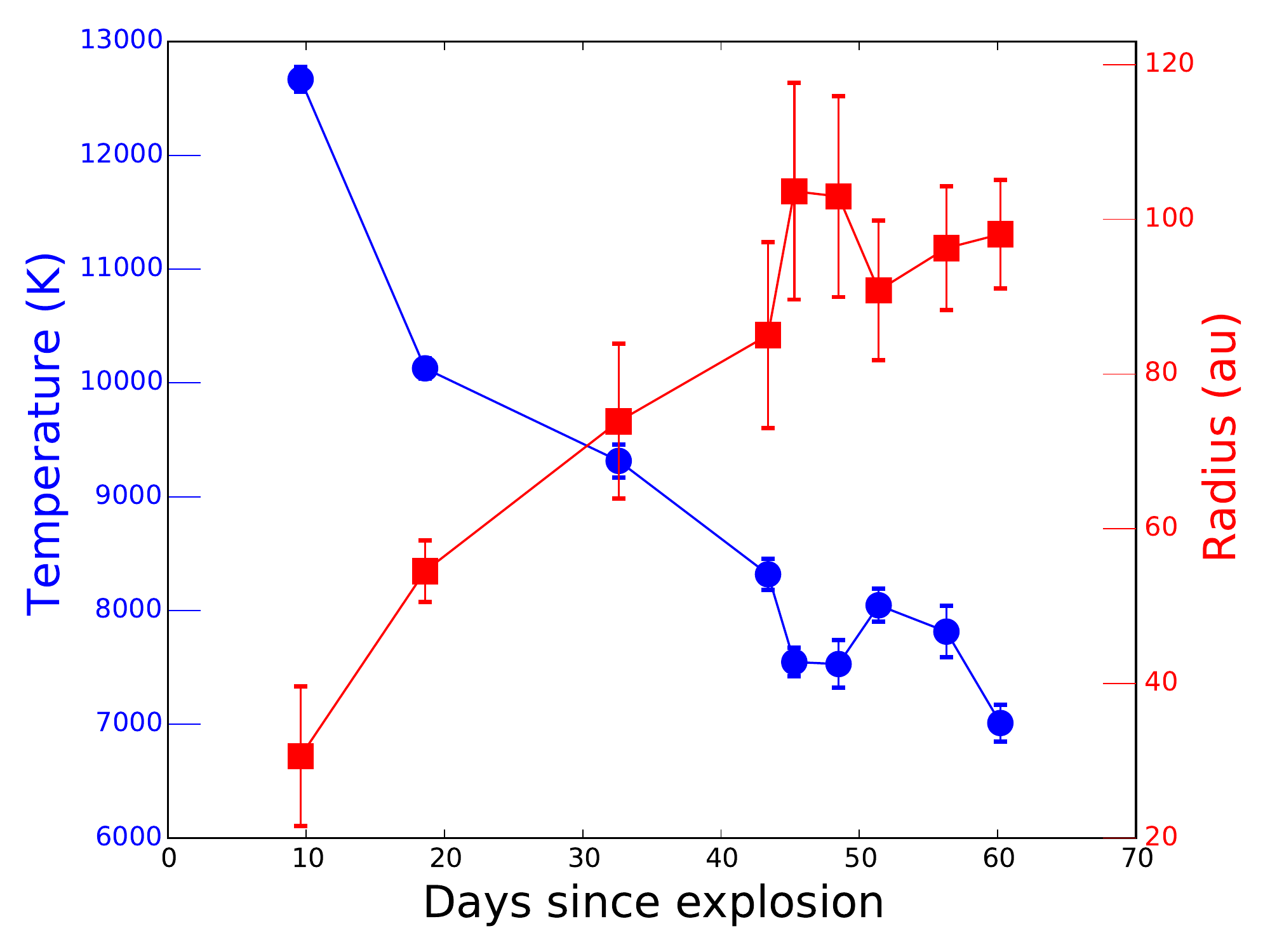}
	\end{center}
	\caption{Radius (right y axis) and temperature (left y axis) evolution of SN 2012ab during the first 2 months after the explosion.}
	\label{fig:rad_temp}
\end{figure}

To this aim we performed black-body fits of spectra  up to 62 d, when we are confident that the black-body approximation holds.
After performing reddening and redshift corrections to the observed spectra, the fits were performed selecting line-free regions in the spectra.  
Typical standard deviations of repeated fits defining different continuum regions are of the order of 200$\degree$K, to which we must add in quadrature the uncertainty on the flux calibration curves and on the galaxy subtraction, where much of the uncertainty resides. We believe that the overall uncertainty on the T determination is of the order of 1000$\degree$K.
Fig.~\ref{fig:rad_temp} illustrates the derived temporal evolution of the temperature.

The black-body fit to the 9.6 d spectrum of SN 2012ab gives a temperature of about 12600$\degree$K
that quick decreases to 10100$\degree$K in less than 10 d. Then the black body temperature decrease progressively more slowly to reach about 8200$\degree$K on 43 d.
From this epoch the large uncertainty due to the galaxy contamination leads to a scatter in the temperature determination which anyway seems to decrease to 7000$\degree$K at two months past explosion. The parent galaxy contamination increases while, at the same time, the emission lines grow in intensity.
The black body fit, therefore, becomes more uncertain.

Having the temperature T and the bolometric luminosity L (cfr. Sect.~\ref{sec:abs_bol}) it is possible to compute the radius of the emitting region using the relation L = 4$\pi$R$^{2}$$\sigma$$T^{4}$.
The radius increases during the first months, and then decreases to reach progressively R$\sim 100$AU at the last epoch (cfr. Fig \ref{fig:rad_temp}, right ordinate). The scatter in the temperature determines the observed dispersion in the radius.

\section{H$\alpha$ evolution}
\label{sec:ha_profile}
The inspection of Figs. \ref{fig:earlyspec} and \ref{fig:latespec} shows a clear evolution of the H$\alpha$ line profile that carries important information about the ejecta and the CSM. 

\begin{figure}
	\begin{center}
		\includegraphics[width=1\linewidth]{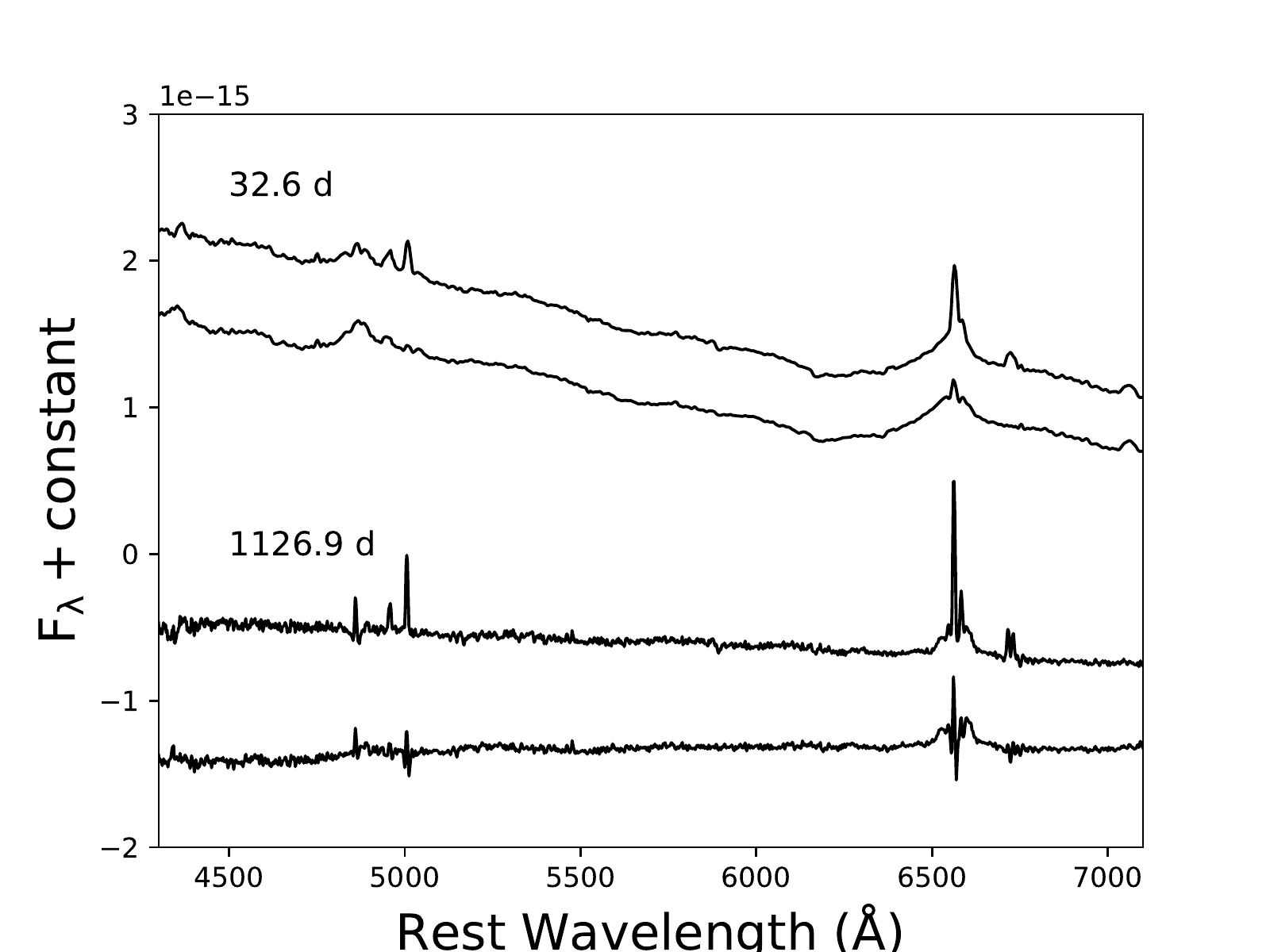} 
	\end{center}
	\caption{Comparison between the reduced spectra of SN 2012ab on 32.6 d and 1126.9 d, with  (bottom) and without (top) the additional removal of the residual galaxy contamination described in the text. At the second epoch (1126.9 d) less-than-perfect subtraction is evident in correspondence of the strongest unresolved lines. Note that the slope of the continuum is unrealistic due to an over-subtraction of the galaxy nucleus spectrum already applied during the normal spectral reduction. After the application of this procedure a broad \Hb\/ emission is visible also in the latest spectrum. }
	\label{fig:totalHalpha}
\end{figure}

Before making a quantitative analysis of the line profile it is important to notice that most of the spectra of SN 2012ab in Figs. \ref{fig:earlyspec} and \ref{fig:latespec} show  unresolved emissions (H$\alpha$, H$\beta$, [O II], [O III], [S II], [N II]) throughout the entire evolution with line ratios typical of H II regions \citep[see also][]{2018MNRAS.475.1104B}. During the data reduction the galaxy background has been removed by subtracting the spectrum of a nearby H II region. Nevertheless, because of the uneven galaxy background and the variable observing conditions, residual unresolved lines remain. In order to understand if the narrow emissions are entirely due to contamination of the host galaxy, we have empirically subtracted to the reduced spectra the host galaxy spectrum retrieved from SDSS  and scaled to remove the [S II] 6717-6731 \AA\ lines. This was done after degrading the resolution of the host galaxy spectra.
This exercise has shown that in general it is possible to remove entirely the narrow-line  contamination of the host galaxy, leaving sometimes negligible residual contamination. Significant exceptions are the earliest 3 spectra for which the cancellation of the [S II] forbidden lines leaves unresolved H$\alpha$ and H$\beta$ emissions at galaxy rest frame. We believe, therefore, that such unresolved emissions are associated to the SN itself (cfr. Table \ref{tab:gauss_log}). This result contradicts the conclusions based on the evolution of the \Ha/\Hb\/ line ratios from \citet[]{2018MNRAS.475.1104B}, who concluded that the narrow lines are always due to host galaxy contamination.
In Fig.~\ref{fig:totalHalpha} we show the results of this procedure at two reference epochs, on 32.6 d, when the SN spectrum is very bright, and on 1126.9 d, when the SN has faded by about 4 magnitudes. 

\begin{figure*}
	\begin{center}
		\includegraphics[scale=0.9]{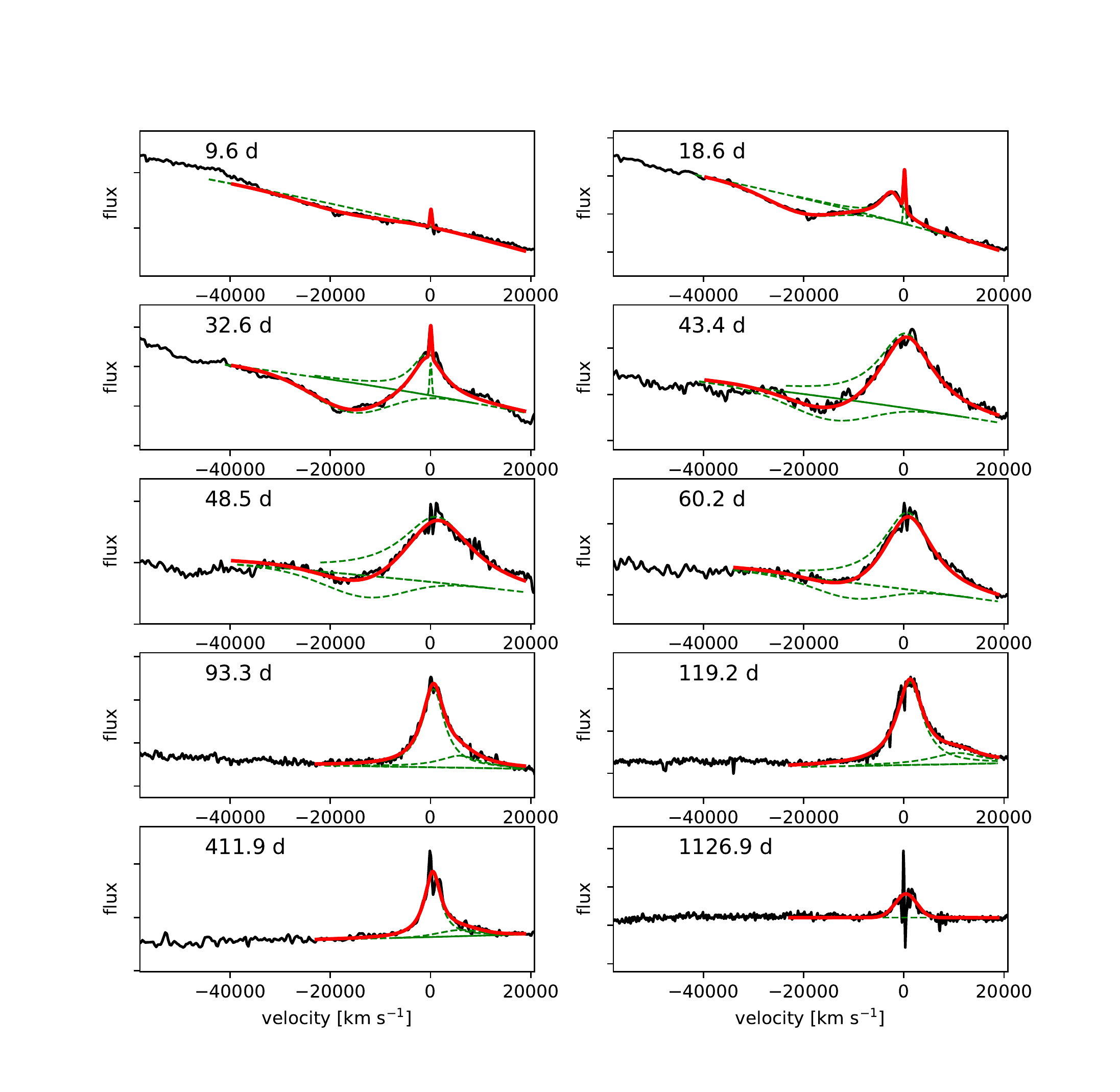}
	\end{center}
	\caption{Best fit (red continuous line) to the observed H$\alpha$ line profiles (black line) and individual components (green dashed) as reported in Table~\ref{tab:gauss_log}. The fit has been performed on background subtracted SN spectra aiming to null the narrow [SII] emissions of the host galaxy, as described in Sect.~\ref{sec:ha_profile}, leaving in some case zero-flux residual (e.g. in the last spectrum) that have not been fitted. Zero velocity corresponds to the \Ha\/ rest position.}
	\label{fig:Halpha_evol}
\end{figure*}

We have, therefore, performed multicomponent line profile fits of the H$\alpha$ emissions after removal of this residual galaxy contamination. The results are displayed in Fig.~\ref{fig:Halpha_evol} where we show the observed spectra along with the individual components used in the fit. Defining and choosing a continuum is very critical for the estimated results. We have selected the continuum in a region external to the line region by at least 50 \AA. The selection of continuum and the consistency of fits were repeated and checked several times. During the early times as several broad lines are present, we had to check the consistency of our results with the fits at adjacent epochs.
 
To describe the line profiles four components are required, of which only three (or less) are required simultaneously for each epoch: an unresolved emission at rest-frame at the earliest epochs, a broad blue-shifted absorption during the first two months, a central emission of intermediate width (IW), centered close to rest- frame during most of the SN evolution, and a red-shifted emission at late times. The fits have been performed using both Gaussian and Lorentzian profiles that produce similar and equally satisfactory results. \cite{2001MNRAS.326.1448C} proposed that in interacting SN spectra a Lorentzian profile of a SN can arise from an intrinsically narrow line that is broadened by collisions with thermal electrons in an opaque shell outside the photosphere. Gaussian profile are generally associated with the emission coming from the P-cygni profile of the ejecta which is in overall a homologous expansion.
Table~\ref{tab:gauss_log} reports the central wavelength of each component, the line flux and the FWHM. For the  absorption also the terminal velocity (at zero intensity) of the blue wing and the expansion velocity derived from the position of the minima are reported. 
The terminal velocity of the red wing of the red emission is also shown in the last column.

In the first spectrum (9.6 d), taken shortly after the rapid rise to maximum, the \Ha\/ line profile shows an unresolved emission at rest-frame, likely due to unshocked CSM associated to the SN (cfr. Sect.~\ref{sec:discussion}), and a broad, shallow absorption centered at about $-$16,000 \kms, extended to $\sim$-$30,000$ \kms. 
Because of its shallowness the line parameters have been derived by using an initial guess of the fit parameters obtained at the second epoch. In Table~\ref{tab:gauss_log} they are reported with the uncertainty mark (:).

Nine days after (18.6 d) the line profile has changed dramatically.  The unresolved emission grows even stronger than before and emerges on the red wing of a much broader (FWHM $\sim5600$ \kms) blue-shifted emission centered at 6511 \AA\/ that, together with a  pronounced absorption, providing a P-Cygni profile to the line. This IW component, centered close to the \Ha\/ rest-wavelength (with some variations with time), will persist during the whole evolution of the SN as in \citet[]{2018MNRAS.475.1104B}. 

There is evidence of an unresolved emission also on 32.6 d, convincing beneath which stands a well developed P-Cygni profile with an IW emission component (FWHM $\sim8200$ \kms) emission that is quickly drifting towards the \Ha\/ rest-frame position. The absorption is still strong and extends to high velocities ($\sim-26600$ \kms).
The minimum of the P-Cygni absorption seems to indicate an high photospheric expansion velocity of 16,600 \kms.

In the following spectrum (43.4 d) the unresolved emission associated to the unshocked CSM has definitively disappeared. 
From this until epoch up to about 60 d (i.e. during the second part of the plateau observed in $R$ and $I$ bands, cfr. Sect.~\ref{sec:lc}) the line profile is well fitted by two components only, a single Lorentzian emission (but a Gaussian works decently as well) in emission and a Gaussian absorption.
It is interesting to note that the center of the absorption, as well as the minimum of the P-Cygni profile and the terminal velocity, change very little with time up to its definitive disappearance at about 60 d.
On the other hand during the same time interval (43.4 d to 60.2 d) in the time window the IW emission reaches its maximum FWHM and luminosity (48.5 d, FWHM $\sim18,000$ \kms, L(\Ha)$=3.36 \times 10^{-13}$ erg cm$^{-2}$ s$^{-1}$) and remains centered to slightly red wavelength ($\sim 6580~$\AA).

On 76.5 d, the IW emission suddenly shrinks to about 6000 \kms, i.e. close to the values required for the second spectrum (18.6 d) and drops in luminosity.
At the same time the line becomes progressively more asymmetric with an extended red wing.
Such red wing can be accounted for a red component centered at about 6800 \AA~ (FWHM $\geq11,000$ \kms), that emerges more clearly only after 119.2 d (cfr. Fig~\ref{fig:Halpha_evol}) persisting up to 817.2 d.  
Starting on 76.5 d, there is therefore a clear indication  of interaction of very fast ejecta ($\sim25,000$ \kms\/ red terminal velocity)  with a receding blob of CSM.
Differently from us, \citet[]{2018MNRAS.475.1104B} have explained the increases on the red wing by introducing a broad component extending from $-15,000$ to $+20,000$ \kms at zero intensity.  We believe that our profile decomposition  is more robust (cfr. our Fig.~\ref{fig:Halpha_evol} with their Figs.~8 and 9) especially from 119 d to 817 d. Nevertheless, despite the different deconvolution, we confirm their finding of an increase of the flux of the receding  (velocity $\ge20,000$ \kms) side of the ejecta between 76 d and 817 d.

The last spectrum 1126.9 d consists of only a symmetric IW component at  the rest wavelength with FWHM $\sim4100$ \kms. Therefore, the ejecta are now interacting only with a more symmetric CSM which is entirely embedding the exploding system. 

\begin{figure}
		\includegraphics[width=1.0\linewidth]{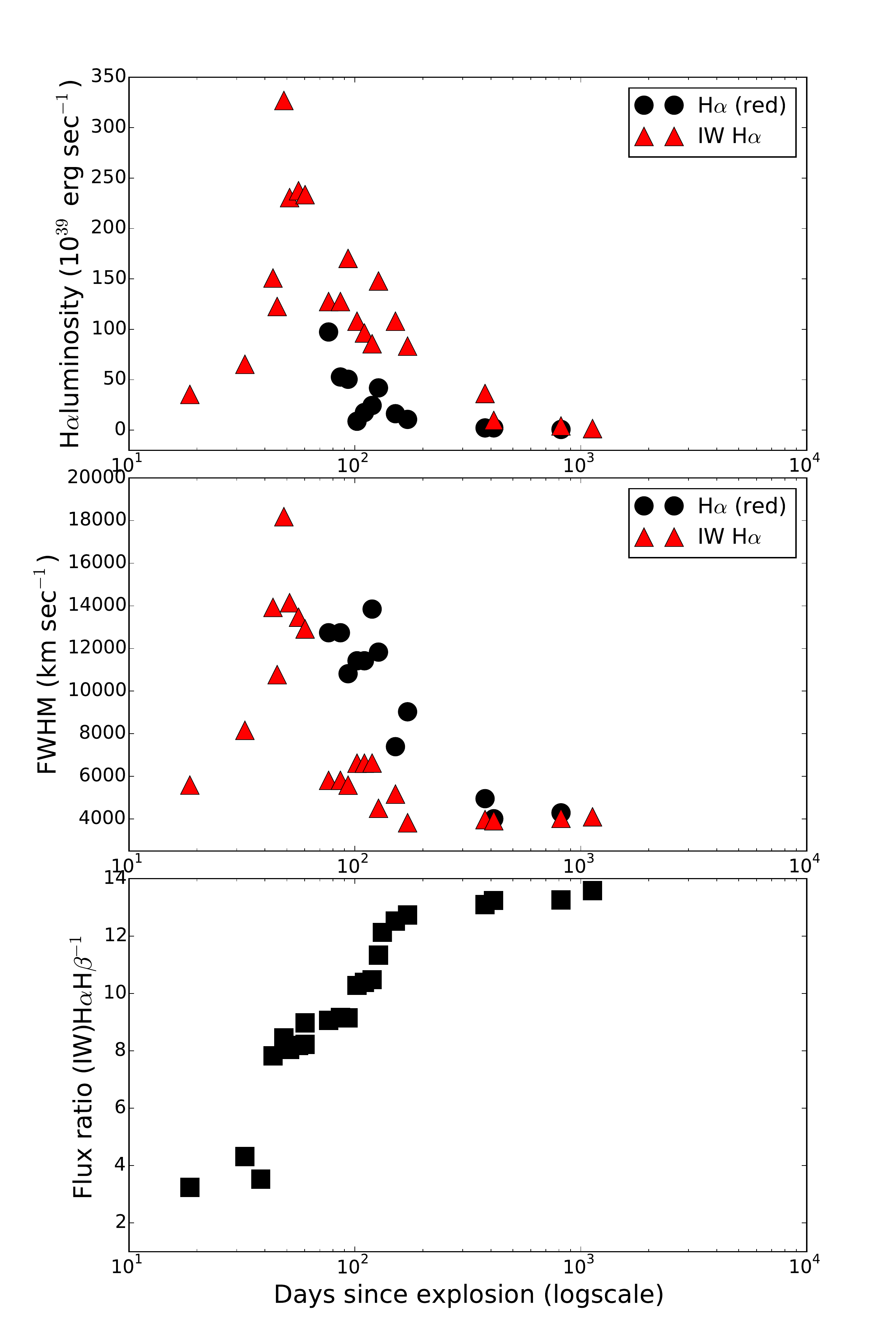}
	\caption{The top panel shows luminosity evolution of the central (black) and red-shifted (red) H$\alpha$ components. The middle panel shows the evolution of the FWHM of the two H$\alpha$ components.
 The bottom panel shows the evolution of the H$\alpha$ to H$\beta$.}
	\label{fig:combinedplot}
\end{figure}

Top and middle panels of Fig \ref{fig:combinedplot} shows the evolution of the \Ha\/ luminosity and FWHM of the various emission components, tabulated in Table \ref{tab:gauss_log}.
The luminosities have been estimated with the distance and reddening discussed in Sect. \ref{sec:galaxy}.
 
From 9.6 d to 48.5 d, the \Ha\/ luminosity of the IW component increases by a factor of $\sim9$. Afterward it is evident a slower, monotonic decline with time that persists up to the latest detection (1127 d). As mentioned above the red component is required only after 76 d. The scatter in the plot gives an idea of the uncertainty of the flux determinations of both components at this epoch ($\pm 20$\%).

\begin{center}
\begin{table*}
\caption{Multicomponent fitting to the H$\alpha$ profile of SN 2012ab. The main parameters of the four components (3 in emission and 1 in absorption) are reported as measured on the observed spectra.}  
\begin{adjustbox}{width=0.99\textwidth,height=0.2\textwidth}
\begin{tabular}{c | c c c | c c c c | c c c | c c c c}
\hline \hline  
Phase  &  \multicolumn{3}{c|}{Narrow emission}                & \multicolumn{4}{c|}{Absorption}                                                      &  \multicolumn{3}{c|}{Central (IW) Emission}           &   \multicolumn{4}{c}{Red Component}        \\
            &   center       & flux                 & FWHM   &   center            & flux  & bleu term. vel.   &  P-Cyg vel.  &      center       & flux                               & FWHM     &      center       & flux                               & FWHM  & red term. vel. \\
 (Days) &   \AA    & erg cm$^{-2}$ s$^{-1}$  & km s$^{-1}$ &   \AA  ~(km s$^{-1}$) & erg cm$^{-2}$ s$^{-1}$  & km s$^{-1}$  &  km s$^{-1}$&   \AA    & erg cm$^{-2}$ s$^{-1}$  & km s$^{-1}$  &  \AA  ~(km s$^{-1}$)   & erg cm$^{-2}$ s$^{-1}$  & km s$^{-1}$ & km s$^{-1}$\\
\hline
	9.6	&	6565	&	1.68E-15	&	<500	&	6206:	(-16305:)	&	-2.38E-14:	&	-30900:	&	---	&	---	&	---		&	---	&	---			&	---	&	---		&	---		\\	
	18.6	&	6566	&	2.97E-15	&	<500	&	6146	(-19048)	&	-3.33E-14	&	-29800	&	---	&	6511	&	3.64E-14	&	5591	&	---			&	---	&	---		&	---		\\	
	32.6	&	6563	&	2.84E-15	&	<500	&	6232	(-15130)	&	-5.77E-14	&	-26600	&	-16600&	6551	&	5.87E-14	&	8195	&	---			&	---	&	---		&	---		\\	
	43.4	&	---	&	---		&	---	&	6262	(-13754)	&	-3.93E-14	&	-23600	&	-15550&	6572	&	1.55E-13	&	13929&	---			&	---	&	---		&	---		\\	
	45.3	&	---	&	---		&	---	&	6233	(-15085)	&	-5.72E-14	&	-23800	&	-16000&	6581	&	1.26E-13	&	10773&	---			&	---	&	---		&	---		\\	
	48.5	&	---	&	---		&	---	&	6291	(-12447)	&	-7.02E-14	&	-23200	&	-16000&	6590	&	3.36E-13	&	18185&	---			&	---	&	---		&	---		\\	%
	51.4	&	---	&	---		&	---	&	6266	(-13581)	&	-5.74E-14	&	-23200	&	-14100&	6585	&	2.37E-13	&	14477&	---			&	---	&	---		&	---		\\	%
	56.3	&	---	&	---		&	---	&	6294	(-12301)	&	-4.52E-14	&	-24100	&	---	&	6580	&	2.44E-13	&	13469&	---			&	---	&	---		&	---		\\	
	60.2	&	---	&	---		&	---	&	6340	(-10212)	&	-4.11E-14	&	-20500	&	---	&	6580	&	2.40E-13	&	12917&	---			&	---	&	---		&	---		\\	
	76.5	&	---	&	---		&	---	&	---				&	---	&			&	---	&	6574	&	1.31E-13	&	5813	&	6666:(4713:)	&	1.00E-13:	&	12739:&	23200:	\\	
	86.3	&	---	&	---		&	---	&	---				&	---	&			&	---	&	6577	&	1.31E-13	&	5813	&	6730:(7634:)	&	5.41E-14:	&	12739:&	24100:	\\	
	93.3	&	---	&	---		&	---	&	---				&	---	&			&	---	&	6576	&	1.75E-13	&	5588	&	6693:(5947:)	&	5.18E-14:	&	10816:&	21900:	\\	
	102.2&	---	&	---		&	---	&	---				&	---	&			&	---	&	6588	&	1.11E-13	&	6619	&	6804:(11016:)	&	9.00E-15:	&	11423:&	22700:	\\	
	110.2&	---	&	---		&	---	&	---				&	---	&			&	---	&	6589	&	9.90E-14	&	6619	&	6804:(11016:)	&	1.79E-14:	&	11423:&	16500:	\\	
	119.2&	---	&	---		&	---	&	---				&	---	&			&	---	&	6590	&	8.80E-14	&	6625	&	6804	(11016)	&	2.51E-14	&	13849&	27200	\\	
	132.2&	---	&	---		&	---	&	---				&	---	&			&	---	&	6582	&	1.52E-13	&	4502	&	6777	(9796)	&	4.30E-14	&	11830&	22700	\\	
	151.2&	---	&	---		&	---	&	---				&	---	&			&	---	&	6582	&	1.11E-13	&	5169	&	6796	(10651)	&	1.67E-14	&	7393	&	18600	\\	
	171.1&	---	&	---		&	---	&	---				&	---	&			&	---	&	6578	&	8.56E-14	&	3823	&	6775	(9691)	&	1.09E-14	&	9030	&	18100	\\	
	376.9&	---	&	---		&	---	&	---				&	---	&			&	---	&	6575	&	3.73E-14	&	3970	&	6774	(9631)	&	2.20E-15	&	4954	&	18600	\\	
	411.9&	---	&	---		&	---	&	---				&	---	&			&	---	&	6572	&	9.84E-15	&	3918	&	6655	(4214)	&	2.23E-15	&	4012	&	9600		\\	
	817.3&	---	&	---		&	---	&	---				&	---	&			&	---	&	6561	&	4.17E-15	&	4031	&	6669	(4854)	&	6.42E-16	&	4269	&	8500		\\	
	1126.9&	---	&	---		&	---	&	---				&	---	&			&	---	&	6572	&	1.50E-15	&	4102	&	---	---	---		&	---	&	---	&	---		\\	
\hline          		                    
\end{tabular}
\label{tab:gauss_log}     
\end{adjustbox}
":" means uncertain determination.
\end{table*}
\end{center}

The central panel of Fig \ref{fig:combinedplot} shows the FWHM evolution of the emission components of H$\alpha$.
The IW component rapidly broadens reaching its maximum FWHM ($\sim18,200$ \kms) at the same time as that of the luminosity peak (top panel). 
Then it monotonically decreases levelling at a FWHM $\sim4000$ \kms~ after 150 d.
Note that during the early phase the central emission is significantly blue-shifted, drifting to the red as  the ejecta becomes optically thin. Then, during the spike in luminosity and FWHM (48.5 d) the line is overall centered at longer wavelengths ($\sim6580$ \AA).
The red component remains broader than the IW until its last detection (on 817 d).
The terminal velocity of the red component is difficult to estimate because of the low signal-to-noise ratio of the late time spectra. However, there is a clear indication that the red wing extends to very high velocities, $v\sim25000$ \kms on 119 d and $v\sim8000$ \kms on 817 d.

Furthermore, although the \Hb\/ flux is not easy to measure because of line blending, we have attempted to measure \Ha\/ to \Hb\/ flux ratio that is shown in the bottom panel of Fig \ref{fig:combinedplot}.
Despite the large uncertainty (nearly $\sim30$\%) a monotonic increase (from $\sim$ 3 to $\sim$ 13 in the time interval ranging from 9 d to 411 d) is clearly observed. During the first month the values are rather constant and similar to Case B  recombination, implying that the H$\alpha$ emitting gas is actually photoionized, like in the early phases of SN 1996al \citep{2016MNRAS.456.3296B}. Already around 45 d the flux ratio rapidly rises to about $8-9$ indicating that probably the gas becomes more collisionally excited. After a sort of plateau, from 90 d we observe that the value monotonically increases to $\sim$ 13 at 170 d, when we believe pure-collisional excitation dominates \citep{1981ApJ...244..780B}. 
	
\section{Discussion}  
\label{sec:discussion}

\paragraph*{AGN,TDE...:} In Sect.~\ref{sec:SNloc} we have shown that our astrometric determination of the SN position is not coincident with the galaxy nucleus (offset $0\arcsec.7$), in agreement with \citet[][]{2018MNRAS.475.1104B}.
These authors have discussed the possibility that SN~2012ab is instead either an AGN or a TDE.
AGN nature was ruled out since AGNs vary stochastically by only a few tenths of magnitude \citep{2012ApJ...753..106M}. From 60 to 120 d, we see a sharp drop of magnitude by about 2 mag in all bands, which makes the AGN scenario unlikely. From a spectroscopic point of view, forbidden lines, such as [O I] 6300, 6364 \AA\/ and [Fe VII] 5721 \AA, are expected in an AGN. Also, the He II 4686 \AA\/ line is sometimes more prominent than the Balmer series in AGNs. These are not detected in the spectra of SN 2012ab. Moreover, high state AGNs show a distinct break in the continuum slope near 5000 \AA~ \citep{2001AJ....122..549V}. Such spectral break is not observed in the case of SN 2012ab.
Similarly, the possibility that SN 2012ab is a TDE was ruled out on the basis of the observed light curve shape and spectral features. 
However, blackbody temperatures estimated for SN 2012ab are lower and evolve faster than those typically observed in TDEs  \citep[see, e.g., PS1-10adi,][]{2017NatAs...1..865K}. The spectra of TDEs are either featureless or show broad emission lines of He I/He II \citep{2012EPJWC..3903001G}. Such lines are not detected in the spectrum of SN 2012ab. \cite{2012EPJWC..3903001G} has provided some tool to distinguish SNe, AGN and TDEs based on the X-ray and UV data. Unfortunately these cannot be applied to SN~2012ab as X-ray and UV observations are not available. 
Based on the data available, we conclude SN~2012ab to be a CCSN with CSM interaction, in agreement with \cite{2018MNRAS.475.1104B}.

\paragraph*{Classification:} The late-time light curves of SN 2012ab is not sustained by the radioactive decay of $^{56}$Co (cfr.  Sect~4.2). Only in the interval between 350 and 440 d the bolometric slope  ($0.78$ \mhund) approached the radioactive dacay but the corresponding $^{56}$Ni mass would be an exceptionally large value.
Overall the light curve of SN 2012ab stays in the average of the strongly interacting SNe of the reference sample discussed in Sect.\ref{sec:comp_sample} both in luminosity and evolution. The spectra do not show the typical evolution of a non-interacting SN II. There is a clear evolution of the normal SED toward lower temperatures in the first two months without change in the spectral lines. The spectra never display the forbidden lines characteristics observed in CCSNe in nebular phase.

\paragraph*{Mass-loss:} We believe that the presence  of the central IW component is a clear indication of the presence of interaction between the ejecta and the CSM, likely originated from the progenitor. We can estimate the mass-loss rate of the progenitor star assuming that the luminosity of the ejecta-CSM interaction is fed by energy at the shock front. The progenitor mass loss rate $\dot{M}$ can be calculated using the relation of  \cite{1994MNRAS.268..173C} :
\begin{equation}
\dot{M}=\frac{2L}{\epsilon}\frac{v_w}{v_{SN}^{3}}
\end{equation}
		
where $\epsilon$ ($<$1) is the efficiency of conversion of the shock's kinetic energy into optical radiation (an uncertain quantity), v$_w$ is the velocity of the pre-explosion stellar  wind, v$_{SN}$ is the velocity of the post-shock shell and L is the bolometric luminosity of SN 2012ab. 
Unfortunately we do not know the velocity of the wind. Our low-resolution spectra just provide an upper limit to the unshocked wind velocity, v$_w\le 500$ \kms, which in turn provide us just an upper limit to the mass-loss.
For the post-shocked shell we assume the characteristic velocity given by the Half width at half maxima (HWHM) of the IW component, that dominates the \Ha\/ emission at all epochs (cfr. Table~\ref{tab:gauss_log}), and for the luminosity the bolometric luminosity computed in Sect. \ref{sec:abs_bol}. Using $\epsilon$ = 1, to be directly comparable to \cite{2018MNRAS.475.1104B}, we have computed the mass-loss at four epochs in which the SN luminosity is dominated by ejecta-CSM interaction, i.e on 76. 119, 411 and 1127 d.
The inferred mass-loss rates are $\dot{M}\leq 0.20$, 0.057, 0.071 and 0.016 \Msy, respectively. These values are compatible with $\dot{M}\sim 0.050$ \Msy derived for 2012 Apr. 17 (79 d) by \cite{2018MNRAS.475.1104B} who used slightly different luminosity and assumed a value v$_w$ = 100 km s$^{-1}$, while we conservatively used v$_w\leq500$ \kms, provided by the observations.
Our derived mass-loss rates are comparable to the values often attributed to some SN IIn, which are of the order of 0.1 \Msy\/ \citep[e.g.][]{2004MNRAS.352.1213C,2007ApJ...656..372G,2012ApJ...744...10K} as observed in some giant eruptions of LBVs, and and instead they appear somewhat larger than those of other SN IIn SN 2005ip: $2-4\times10^{-4}$ \Msy, \citet{2009ApJ...695.1334S}, SN 2017hcc: 0.12 \Msy, \citet{2019MNRAS.488.3089K} and it is also much larger than the typical values of RSG and yellow hypergiants $10^{-4}-10^{-3}$ \Msy \citep{2014MNRAS.438.1191S}, and quiescent winds of LBV (10$^{-5}-10^{-4}$ \Msy, \cite{2018A&A...619A..54V}. 

\cite{2018MNRAS.475.1104B} have provided a general physical picture for SN 2012ab according to which the SN interacts with a largely asymmetric CSM. In particular, they claimed to observe firstly ejecta-CSM interaction occurring on the side to the observer and, later on, on the opposite side. They proposed that the CSM lay on an equatorial plane, seen at an intermediate inclination, and interpret this as evidence of non-axisymmetric mass loss in an eccentric binary system.
Their evidence was based on the lack of normal P-Cygni features, and on the presence and timing of blue- and red-shifted emission features.

\paragraph*{Physical scenario:} Based on our analysis of Sect.~\ref{sec:ha_profile} we propose some variation to this picture.
Our new analysis of the HET spectra shows that unresolved emission lines associated to the SN are present up to 32 d. They are attributed to unshocked slow-moving CSM, ionised by the X-UV flash shock break-out in close analogy to that observed in SN 1996al \citep{2016MNRAS.456.3296B}.  There is no evidence of significant narrow absorption associated to this gas even in the  MMT/BC high resolution spectrum obtained on 2012 March 1 (32 d) \citep[cfr.][]{2018MNRAS.475.1104B} and is, therefore, either fully ionized and/or aligned out of the line of sight. 

The narrow emission is no longer visible in our spectrum taken on 43.6 d, therefore  by this epoch the unshocked gas either recombined quickly was swept away. The external boundary of the unshocked gas is R$_{CSM}^{out}\geq(v_{ej}^{term}-v_{gas})\times 32.6 d\sim 10^{16}$ cm from the exploding star. It is, therefore, reasonable to conclude that significant interaction  between the ejecta and the CSM takes place soon after the burst.

In previous Sections we have shown that on 9.6 d very broad and shallow absorption can be recognized at wavelengths shorter than the \Ha\/ rest position.
Subsequent spectra taken between 19 d and 60 d show more clear evidence of broad absorptions (Fig.~\ref{fig:Halpha_evol}), extending to very high  terminal  velocities (v$_{term}\sim30,000$ \kms at zero intensity, Table~\ref{tab:gauss_log}) giving to \Ha\/ a very broad P-Cygni profile.
There is, therefore, evidence of a photosphere in an expanding envelope that at early phases appears to be optically thick and obscures the receding ejecta.
We note that during the period in which the broad absorption is visible, the width of the central emission is unusually broad (FWHM up to 18,000 \kms). 

At this stage, the observed spectrum is probably a combination of radiation from an high-energy, H-rich fast expanding envelope of the SN diluted by a continuum emission from the ejecta-CSM interaction. Anyway, the line profile does not provide evidence for an asymmetric interaction stronger on the approaching side of the ejecta at early epoch, as suggested by \citet{2018MNRAS.475.1104B}. 

The minimum of the P-Cygni absorptions, in particular, those of Fe II lines, is typically used to determine the photospheric velocity. Lacking the Fe II absorptions, we can get an upper limit from minimum of the \Ha\/ which, being stronger, forms at higher velocities \citep{1990sjws.conf..149J}. 
On 32 d the \Ha\/ absorption is still very broad and the minimum poorly defined. But anyway, using v$_{ph}(32.6 d)=16,600$ \kms (Table~\ref{tab:gauss_log}), we get a value of the radius of the photosphere four times larger than R$_{BB}$ (cfr. Sect.~\ref{sec:rad_temp}). This is formally compatible, considering the large uncertainties in the photospheric velocity and in the T determinations, but probably it is an indication that some deviation from spherical symmetry is present.  
Indeed, the polarimetric observations  have shown that the geometry of the system from 54 d to 78 d is heavily aspherical \citep{2018MNRAS.475.1104B}. From the similarity with SN 2010jl they suggest a pole-to-equator density ratio of $\sim 2.6$, in analogy to \cite{2015MNRAS.449.4304D}. 
The evidence of asymmetry, the measurements of exceptionally high expansion velocities and the presence of unshocked gas not aligned along the line-of-sight are suggestive of an highly energetic and asymmetric explosion observed along the jet emission \citep[cfr.][]{2009A&A...504..945T,2019A&A...628A..93P}.

Soon after 32 d the photosphere becomes optically thin and the red wings of \Ha\/ rapidly grows in intensity.
Overall the line starts to be skewed to the red and the line centroid is slightly redshifted. 
Such red drift can indeed signal the onset of interaction on the receding side of the fast ejecta, supporting aspherical geometry in the emitting material as suggested by \citet{2018MNRAS.475.1104B}.

By 76 d the absorption has completely disappeared and the central (IW) emission shrunken to FWHM $\sim6000$ \kms. The narrowness of the emission allows the identification of a red wing of the IW component. We note that the terminal velocity of the red component, i.e. the fastest expansion velocity of the receding part of the ejecta in direction opposite to the observer, is comparable to the terminal velocity observed for the absorption, i.e. the fastest ejecta velocity in the direction of the observer. This on one side strengthens the identification of the broad absorption at early phases and the determination of its terminal velocity, and also the hypothesis that the receding ejecta travelled undisturbed until the collision marked by the onset of the red component. The very high measured velocities, of the order of $25,000-30.000$ \kms, support the idea that we are observing along the direction of a jet-like ejecta expanding until this epoch in a region devoid of interaction with CSM. 

Moreover, the observed terminal velocities also suggest that the emission arises in the ejecta shocked by the reverse shock. With time (starting fom 127 d) the terminal velocities decrease because the reverse shock receded into slower layers. 
Despite the introduction of this additional red component, the central (IW) component maintains a slightly redshifted wavelength position (corresponding to $v\sim550$ \kms). The IW component instead may likely arises from shocked CSM in a cocoon enclosing the jet.

After 300 d there is a clear increase in the slopes of light curves at all wavelengths that mimics for few months (up to about 440 d) that of $^{56}$Co decay (cfr. Table~\ref{tab:comp_decay_rate}). The overall spectral features remain unchanged. The \Ha\/ profile still requires the red component, but its FWHM appears significantly narrowed and the terminal velocity halves quickly. We read this as evidence that the shock with the receding material is temporarily reduced. 

But again the light curves flatten after 440 d and remain so for at least 2 more years. In the meantime the SN has faded considerably and it is hard to disentangle its luminosity from that of the galaxy nucleus. However, the spectral observations have detected the fading IW component (only) at least up to 3.5yr past the explosion
\citep[cfr. Table~\ref{tab:gauss_log} and ][]{2018MNRAS.475.1104B} confirming, therefore, that the interaction was still going on with a rather symmetric CSM.

\begin{figure}
	\begin{center}
		\includegraphics[width=1.0\linewidth]{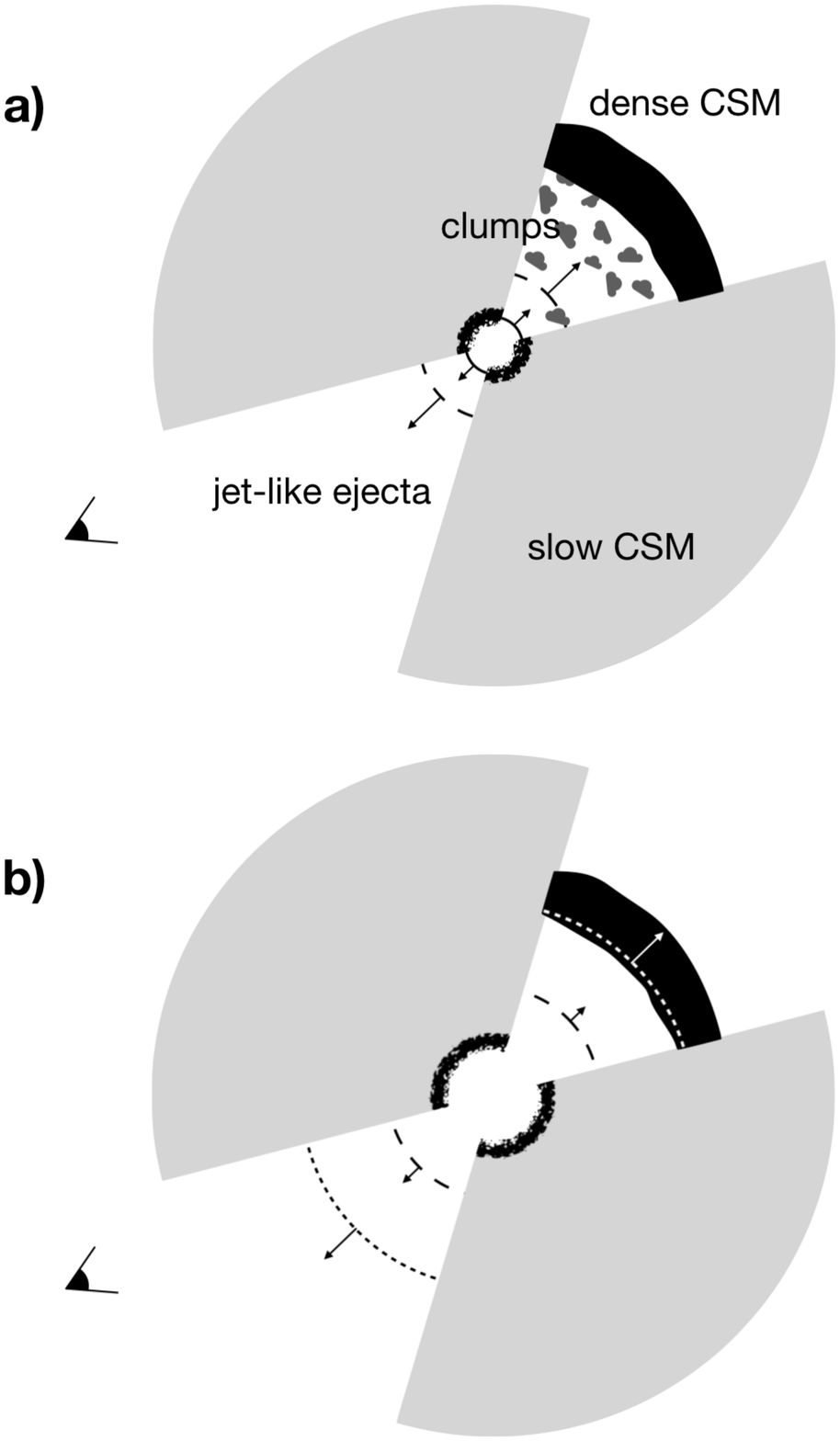}
	\end{center}
	\caption{Sketch of the configuration of SN 2012ab and its CSM at about (a) 40 d and about (b) 80 d past explosion. The observer is placed in the bottom left corner of both panels.}
	\label{fig:sketch}
\end{figure}

During the whole evolution the line profile does not show evidence of dust formation  inside the ejecta as observed in SN 1987A and other CCSNe \citep{2009ApJ...695.1334S,2009ApJ...691..650F,2018ApJ...859...66S}. Fig.~ \ref{fig:sketch} reproduces schematically the above interaction scenario at two epochs.

\section{Conclusions}
\label{sec:conclusions}
This paper presents the photometric and spectroscopic observations of the luminous Type IIn SN 2012ab. 
The main results are summarized as follows:
\begin{enumerate}
\item The {\it UBVRI}~ light curves underwent a multi-stage evolution with  synchronous changes. The rise to maximum was very steep ($\geq3.4$ mag in 4 d in the R band). Afterward, while $R$ and $I$ bands showed a sort of flat plateau lasting about 60-70 d, followed by a steeper decline up to about 100d. In the same, $U$ band declined at constant rate.  Starting from 4-5 months after maximum  the bolometric light curves of non-interacting Type II SNe follow the radioactive decay of $^{56}$Co $\rightarrow$ $^{56}$Fe  \citep[0.98 \mhund,][]{1990AJ....100..771T}. In the same period $B$ and $V$ bands showed a sloping plateau steeper at shorter wavelengths. The bolometric light curve of SN 2012ab remain always flatter than this value.
The peak absolute magnitude is M$_R$ = $-$19.39 mag and the pseudo-bolometric luminosity at maximum is L$\sim$8.3$\times$10$^{42}$ erg s$^{-1}$, making SN 2012ab  brighter than non-interacting SN II \citep{2009Natur.459..674V}. We thus propose that both  radioactive decay of $^{56}$Co and CSM interaction contribute to the high SN luminosity observed at late stages.

\item At all epochs the spectra appear contaminated by unresolved emission lines typical of H II regions. However, our analysis (Sect.~\ref{sec:ha_profile}) shows that part of the observed unresolved flux of \Ha\/ and \Hb\/  observed during the first month is due to slowly expanding, ionized gas in close proximity of the exploding star, while at later epochs the the observed unresolved lines are due exclusively to galaxy contamination.

\item At the epoch of the first spectral observation (9.6 d after the burst), the spectrum of SN 2012ab is characterized by a black-body continuum of T$=12600$~$\degree$K like normal CCSNe, with shallow undulations, possibly indication of very fast absorbing material, plus the above mentioned unresolved emission lines.
In the following spectra, broad H Balmer lines appear with P-Cygni profile due to a very fast expanding envelope (v$_{term}^{abs}\sim30,000$ \kms; Sect.~\ref{sec:ha_profile}). Obscuration by the opaque photosphere produces significant blueshift of the emission component. With time the photosphere shrinks in velocity coordinates, becomes transparent and the obscuration rapidly disappears.
The broad P-Cygni absorptions and the corresponding broad emission component (FWHM $\sim18,000$ \kms) from the scattering envelope are visible for about 2 months. With the decrease of the strength of the broad emission, we identify an IW component (FWHM $\sim6000$ \kms)  with a Gaussian (or Lorentzian) profile that characterizes the line shapes for over 3 years. This is attributed to ejecta-CSM interaction with a globally symmetric CSM. 

\item Starting already on 43 d, the \Ha\/ emission component is slightly moved to the red with respect to the rest-wavelength. With the progressive reduction of the emission FWHM a red component stands out on the red wing of an overall symmetric central emission of intermediate width (IW) that we attribute to the onset of interaction of the receding ejecta with CSM. 
The red component maintains the same central position and FWHM for about 100 d. The (receding) terminal velocity of the red component (v$_{term}^{em}\geq25000$ \kms) is similar to the (approaching) terminal velocity of the absorption observed at early time, an indication that (at least a substantial part of) the ejecta in direction opposite to the observed travelled undisturbed until this new interaction.
	
\item The long-lasting presence of very fast material along the line of sight, that we observe approaching and producing the broad absorption for the first two months, and then receding, and interacting with the CSM, on the receding side from several months more, is suggestive of the fact that we observe the SN along the axis of a jet-like ejection either devoid of CSM in the innermost regions or may be uninterrupted by nearby CSM clumps. Of course, we suggest that there may be clumps of CSM interaction occurring in the approaching region but there are gaps where the SN ejecta are expanding uninterrupted, producing a broad P-Cygni profile and an IW component on the blueshifted side (of variable strength depending on how much CSM material is present). Alternatively, the simultaneous presence of an IW component (FWHM $\sim$ 4000 - 6000 km s$^{-1}$), on the other side, tells that ejecta-CSM interaction with a more symmetric CSM component takes place on the sides of the jet, in addition to the asymmetric one. A sketch of a possible configuration visualizing the findings above is shown in Fig.~\ref{fig:sketch}. Thus, to summarise the Fig.~\ref{fig:sketch}, we say that the long-lasting presence of very fast material along the line of sight, that we observe 1) approaching and producing the broad absorption during the first two months, and then 2) receding, and interacting with the CSM, for several months more, suggests that we observe the SN along the axis of a jet-like ejection in a region either devoid of CSM or may be unhindered by the nearby CSM clumps.

\item We measured the mass-loss rate of the progenitor star to be $\leq0.20-0.016$ \Msy at epochs between 76 d and 1127 d, similar to the value derived by \citet{2018MNRAS.475.1104B}.
This value is consistent with of giant eruptions of LBV. 

\end{enumerate}
	
Therefore, SN 2012ab appears as the outcome of an energetic core-collapse explosion in a dense structured CSM produced by the massive progenitor in the last stages of its evolution via strong mass-loss. However, differently from \citet{2018MNRAS.475.1104B} we do not find evidence of structures in the CSM that can be associated with the progenitor being in an eccentric binary system. 
On the other hand, our analysis supports the presence of both interaction of the ejecta with  a relatively symmetric CSM with abundant nearby clumps as depicted in Fig.~\ref{fig:sketch} from early phase, and the late collision of the receding high-speed (v$\sim25000$ \kms) ejecta with other asymmetric structures of the CSM. 
This, together with the similarly high terminal velocity of the absorptions, suggests that we are observing the interaction from a line of sight aligned with a jet-like ejection in a region devoid of interaction with CSM close to the progenitor star.









{\bf Data Availability Statement:} The data underlying this article will be shared on reasonable request to the corresponding author.

\noindent

\section*{Acknowledgements}

N.E.R. acknowledges support from the Spanish MICINN grant ESP2017-82674-R and FEDER funds. 
SB, LT and MT are partially supported by the PRIN-INAF 2016 with the project {\it Towards the SKA and CTA era: discovery, localisation, and physics of transient sources} (P.I. M. Giroletti). This work is partially based on observations of the European supernova collaboration involved in the ESO-NTT and TNG large programmes led by Stefano Benetti. This work is partially based on observations made with the ESO Telescopes at the La Silla and Paranal Observatories under programme IDs 184.D-1140, 184.D-1152; the Italian Telescopio Nazionale Galileo (TNG) operated by the Fundaci\'on Galileo Galilei of the INAF (Istituto Nazionale di Astrofisica) at the Spanish Observatorio del Roque de los Muchachos of the Instituto de Astrof\'isica de Canarias under the programme ID A25TAC49; the Nordic Optical Telescope, operated by the Nordic Optical Telescope Scientific Association at the Observatorio del Roque de los Muchachos, La Palma, Spain, of the Instituto de Astrofísica de Canarias;  the 1.82m Copernico telescope (Asiago, Italy) of the INAF - Osservatorio Astronomico di Padova; Asiago Observatory; the 1.22 m Galileo telescope of Dipartimento di Fisica e Astronomia (Universit\'a di Padova); the William Herschel Telescope, operated on the island of La Palma by the Isaac Newton Group in the Spanish Observatorio del Roque de los Muchachos of the Instituto de Astrof\'isica a de Canarias;  the 2.2-m Telescope of the Centro Astron\'omico Hispano Alem\'an (Calar Alto, Spain); the Gran Telescopio Canarias (GTC), installed at the Spanish Observatorio del Roque de los Muchachos of the Instituto de Astrof\'isica de Canarias, in the island of La Palma.  We thank the observers at ARIES to support these observations.


\bibliographystyle{mnras}
\bibliography{refag}

\bsp	
\label{lastpage}
\end{document}